\documentclass[sigconf]{acmart}

\usepackage{algorithmic}
\usepackage{textcomp}
\usepackage{xcolor}
\usepackage[utf8]{inputenc}
\usepackage{multirow}
\usepackage{subfig}
\usepackage{algorithm}

\setcopyright{licensedusgovmixed}
\copyrightyear{2024}
\acmYear{2024}
\acmDOI{10.1145/3650200.3656608}

\acmConference[ICS '24]{Proceedings of the 38th ACM International Conference on Supercomputing}{June 4--7, 2024}{Kyoto, Japan}
\acmBooktitle{Proceedings of the 38th ACM International Conference on Supercomputing (ICS '24), June 4--7, 2024, Kyoto, Japan}
\acmISBN{979-8-4007-0610-3/24/06}

\begin{document}

\title{Shared Virtual Memory: Its Design and Performance Implications for Diverse Applications}

\author{Bennett Cooper}
\email{bwc@clemson.edu}
\orcid{0009-0003-0375-9418}
\affiliation{%
  \institution{Clemson University}
  \city{Clemson}
  \state{South Carolina}
  \country{USA}
}

\author{Thomas R. W. Scogland}
\email{scogland1@llnl.gov}
\orcid{0000-0001-7234-5743}
\affiliation{%
  \institution{Lawrence Livermore National Lab}
  \city{Livermore}
  \state{California}
  \country{USA}
}

\author{Rong Ge}
\email{rge@clemson.edu}
\orcid{0000-0002-2218-3675}
\affiliation{%
  \institution{Clemson University}
  \city{Clemson}
  \state{South Carolina}
  \country{USA}
}

\begin{abstract}
Discrete GPU accelerators, while providing  massive computing power for supercomputers and data centers, have their separate memory domain. Explicit memory management across device and host domains in programming is tedious and error-prone. To improve programming portability and productivity,  Unified Memory (UM) integrates GPU memory into the host virtual memory systems, and provides transparent data migration between them and GPU memory oversubscription. Nevertheless, current UM technologies cause significant performance loss for applications.  With AMD GPUs increasingly being integrated into the world's leading supercomputers, it is necessary to understand their Shared Virtual Memory (SVM) and mitigate the performance impacts. In this work, we delve into the SVM design,  examine its interactions with applications' data accesses at fine granularity, and quantitatively analyze its performance effects on various applications and identify the performance bottlenecks. Our research reveals that SVM employs an aggressive prefetching strategy for demand paging. This prefetching is efficient when GPU memory is not oversubscribed.  However, in tandem with the eviction policy, it causes excessive thrashing and performance degradation for certain applications under oversubscription. We discuss SVM-aware algorithms and SVM design changes to mitigate the performance impacts. To the best of our knowledge, this work is the first in-depth and comprehensive study for SVM technologies.
\end{abstract}



\maketitle

\section{Introduction}
\label{sec:intro}
Discrete GPU accelerators, with their massive parallel processing capabilities and energy efficiency,  are crucial for providing the computing power of today's HPC systems and data centers. They enable significant advancements in areas such as climate modeling~\cite{tal2022}, bio-informatics~\cite{pham23}, and AI. As of today, 90\% of top 10 and more than 35\% of top 500 supercomputers are accelerated by discrete GPUs~\cite{top500}, providing about 50\% of the performance share of top 500 supercomputers.
They have also become the predominant hardware for deep learning and large language models (LLMs) training. For example, the BLOOM (176B parameters)  is trained over 1 million GPU hours using BigScience infrastructure~\cite{bloom}, and the GPT-3 (175B parameters)  is estimated to be over several million GPU hours~\cite{brown2020}.

Discrete GPU accelerators have their own separate memory domains from the host memory domain. Programmers must invoke memory copy functions and ensure the data residing in the GPU doesn't exceed GPU memory capacity. Explicit memory management and data movement across domains are laborious and error-prone, especially for memory-demanding workloads where the memory footprint exceeds GPU memory. It is increasingly important to relieve programmers of such tasks and make GPU programming more productive and portable, as deep learning models and social networks are increasingly larger~\cite{bloom,brown2020,emogi,faimgraph}, and scientific workloads are more data-intensive~\cite{tal2022,pham23}.  

Unified Memory (UM) integrates GPU memory into the host virtual memory systems and transparently migrates data between them. Additionally, UM supports \textit{GPU memory oversubscription}, i.e., GPU kernels access more data than the GPU memory can hold, significantly enhancing programming portability and productivity for memory-demanding workloads. UM technologies have been adopted by HPC frameworks such as Raja~\cite{raja}, Kokkos~\cite{kokkos}, and Trilinos~\cite{trilinos} for writing portable applications on today's and future's major HPC platforms, and by deep learning frameworks~\cite{Choi2022ImprovingOG,Jung2023DeepUM,Min2021PyTorchDirectEG}. However, even with active research and improvement by vendors and research community~\cite{oneAPI,opencl,rocm,nvidia-hmm}, current UM technologies cause significant, or even prohibitive, performance degradation~\cite{landaverdeHPEC,Knap2019-fl,Yu2019b}.

To bridge the performance gap between explicit memory management and Unified Memory (UM), a deep understanding of UM's design and identification of performance bottlenecks are crucial. NVIDIA GPUs have been the primary choice for accelerators in supercomputers and data centers, sparking significant research interest in NVIDIA's Unified Virtual Memory (UVM). Researchers have examined UVM's design, its impact on application performance, and proposed optimization techniques~\cite{ganguly19,li19,chang21}. In recent years, AMD GPUs have seen a notable growth in adoption, powering 7 of the 10 most energy-efficient supercomputers. Both the fastest supercomputer, Frontier, and the upcoming leader, El Capitan~\cite{llnl-elcap}, utilize AMD GPUs for acceleration. However, AMD's UM technology, Shared Virtual Memory (SVM), hasn't received much research attention.

SVM has a distinct design from UVM, and the insights derived for UVM may not be directly applicable to SVM. While both are implemented as software drivers mirroring page tables on both the host and the device, UVM is a technology developed for NVIDIA's specific hardware and drivers for performance, optimization, and efficiency. In contrast, SVM interfaces with the Linux kernel's Heterogeneous Memory Management (HMM)~\cite{linux-hmm, hmm}, which is designed for broader hardware compatibility and integration. While HMM is still in development, its unified framework is poised to greatly simplify driver development and enhance application portability.

In this work, we delve into the design of the shared Virtual Memory (SVM) and examine its impact on the performance across a variety of applications. We investigate its UM management strategy, architecture and components, page fault handling and data migration/eviction in demand paging.  We further quantify its  overhead and the overall cost, and identify its performance bottlenecks and its variations with applications' data access. Using the fine-grain fault and migration profiles, we classify the applications and their access patterns, and reason the root causes of performance bottlenecks.

We reveal that SVM manages the unified memory by ranges, i.e., a range is a number (typically large) of contiguous pages. With demand paging, a single fault can trigger an entire range migration, which in turn requires a range eviction if GPU memory is oversubscribed.  \textit{This management strategy is equivalent to the most aggressive prefetching}. We find that the current SVM design is beneficial if the GPU is not oversubscribed, but otherwise causes excessive thrashing and performance degradation for applications with certain temporal and spatial access patterns. Due to limited information available, the eviction policy may evict the most intensely reused data, further exacerbating thrashing. Our quantitative analyses uncover that severe thrashing not only increases the eviction-to-migration ratio, but more seriously increases the number of migrations by orders of magnitude for certain applications. We establish that SVM-aware algorithm designs can significantly improve performance, and discuss possible augmentations to SVM design that could benefit broader applications.  

We make the following main contributions: 
\begin{itemize}
    \item We reveal the SVM design and range migration/eviction in demand paging, and  quantitatively analyze the UM management overhead and the overall costs at fine granularity.  We identify the performance bottlenecks, their significant increases under oversubscription, and  their variations across a variety of applications. 
    \item We unveil the migration and eviction profiles and fault behaviors of diverse applications resulting from the interaction between their memory access and SVM. These profiles are indicative of  performance and expose premature evictions and severe thrashing in applications with intensive data reuse or distributed data accesses.  
    \item We investigate potential benefits from SVM-aware application algorithm design using case studies. We show that SVM-awareness empowers us to mitigate performance bottlenecks and improve performance by up to orders of magnitude. We further discuss potential augmentations to SVM designs to benefit a broader range of applications. 
    \item To the best of our knowledge, this work is the first in-depth and comprehensive study of SVM technology. As various accelerators and devices are expected to interface with Linux HMM, this work may shed light on optimal driver-specific designs and library implementations for target applications. 
\end{itemize}
\section{SVM Design and Architecture}
\label{sec:design}
Unified Memory (UM) integrates the device memory domain into the host's virtual memory system, providing a shared address space for host processes and GPU kernels. UM transparently migrates data between the two memory domains, and keeps track of the physical memory locations on page tables, eliminating the need for programmers to copy explicitly. UM supports oversubscription by evicting old pages from GPU memory before migrating new ones. 

Shared Virtual Memory (SVM)  driver interfaces with Heterogeneous Memory Management (HMM) to interact with host page tables. HMM is expected to be the standard Linux interface for various driver modules. Unlike the one-way communication from the older drivers to the kernel, HMM creates a truly unified memory by preventing the kernel from moving pages without alerting the driver and causing many edge cases of failures. 

\begin{figure}
\centering
\includegraphics[width=0.8\linewidth]{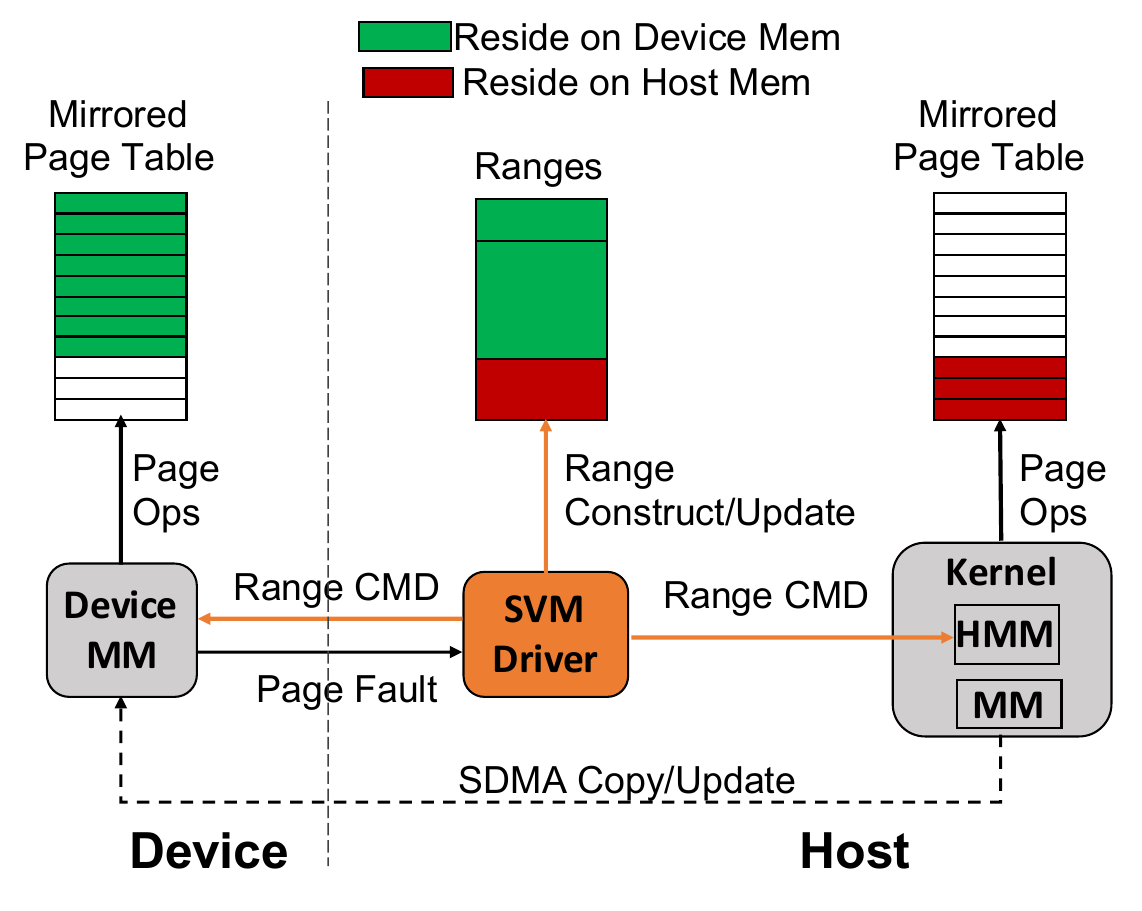}
\vspace{-10pt}
\caption{SVM manages the UM space by ranges, rather than pages in host and device memory domains}
\vspace{-10pt}
\label{fig:HMMinterface}
\Description{Depiction of SVM interplay with host and device memory management.}
\end{figure}

\begin{figure}
\centering
\includegraphics[width=0.7\linewidth]{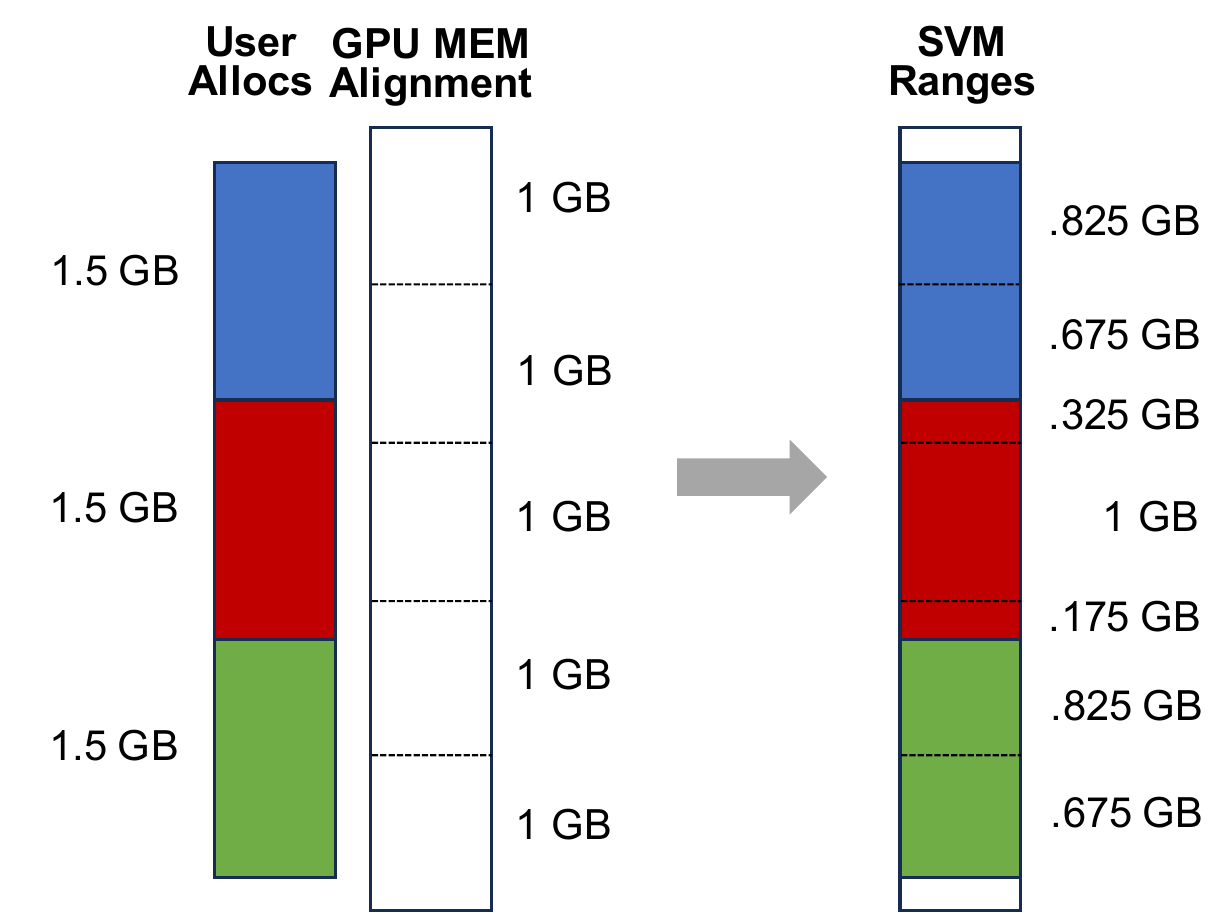}
\vspace{-5pt}
\caption{Example range creation for three 1.5 GB allocations}
\vspace{-18pt}
\label{fig:exampleconstruction}
\Description{Example construction of SVM ranges from three allocations.}
\end{figure}

 In this section, we detail SVM's design and architecture. Our experimental platform is one node from the LLNL Tioga supercomputer~\cite{llnl-tioga}; the node architecture matches those of the Frontier supercomputer~\cite{frontier,amdRetro,mi200-arch}. A node consists of a 64-core AMD 7A53 EPYC CPU, 512 GB DDR4 host memory, and four AMD Instinct MI250X discrete GPUs connected to the host by 36GB/s bidirectional Infinity Fabric. Each MI250X has two GPU compute dies connected by 200GB/s bidirectional Infinity Fabric, each with 64 GB HBM2E memory. Tioga uses the Tri-Lab Operating System Stack~\cite{toss} version 4 \& amdgpu version 6.3.6, with 1 GB GPU memory alignment in SVM. We use ROCM version 5.4.0. The experimental results presented in this work only use one GPU compute die.
 
 \subsection{Ranges as SVM Management Units}
 
Even though both host and device memory domains manage their own address space in pages, SVM manages the unified memory in \textbf{ranges}, as shown in Figure~\ref{fig:HMMinterface}. Each SVM range is defined by a start address and an end address and may comprise a substantial number of contiguous virtual pages. The management operations include allocation and deallocation, migration, and eviction. 

Upon the receipt of a managed memory allocation from the runtime, SVM constructs the ranges based on GPU memory alignment and the allocation's size. GPU memory alignment is determined by its capacity, i.e., $\lfloor \frac{\text{capacity}}{32} \rfloor$ rounded down to the nearest power of two, and should be minimally 2 MB. For example, if a GPU has 48 GB available for SVM managed memory, then the alignment is 1 GB. In addition, the ranges must be aligned to allocation boundaries.  With this range construction, an allocation should comprise multiple ranges if it is large or across alignments.  

Figure~\ref{fig:exampleconstruction} depicts the range construction for an application with three 1.5 GB allocations on a GPU aligned by 1 GB. The application mimics matrix multiply. SVM constructs 7 ranges  of  varying  sizes for this application, with the smallest range at 175 MB and the largest at 1 GB. 

SVM receives faults at the page level from the device but services with data migration at the range level. While a range may consist of up to 256K pages,  it only requires the servicing of a single page fault to trigger the migration of the entire range, with the remaining faults being dismissible. Thus, a received fault undergoes an initial examination to determine if it is serviceable. 

A fault is considered serviceable if it is recent and not duplicate. Recent faults are those with  timestamps falling within the specified timeout period. Old unsatisfied faults would be replayed by the GPU to generate recent faults. A recent fault is considered a duplicate if it originates from the same page or range as a recent range migration. Duplicate faults typically dominate, representing 97-99\% of the total faults generated by a kernel. They can arise from various sources, including the same thread block processing data with spatial locality, different thread blocks processing the same data, and the same thread block processing data with temporal locality. In practice, applications with a high degree of duplication of faults can perform efficiently by consolidating them into a single range transfer.

\subsection{Page-Level Fault and Range-Level Migration}
\begin{figure}
\centering
\includegraphics[width=\linewidth]{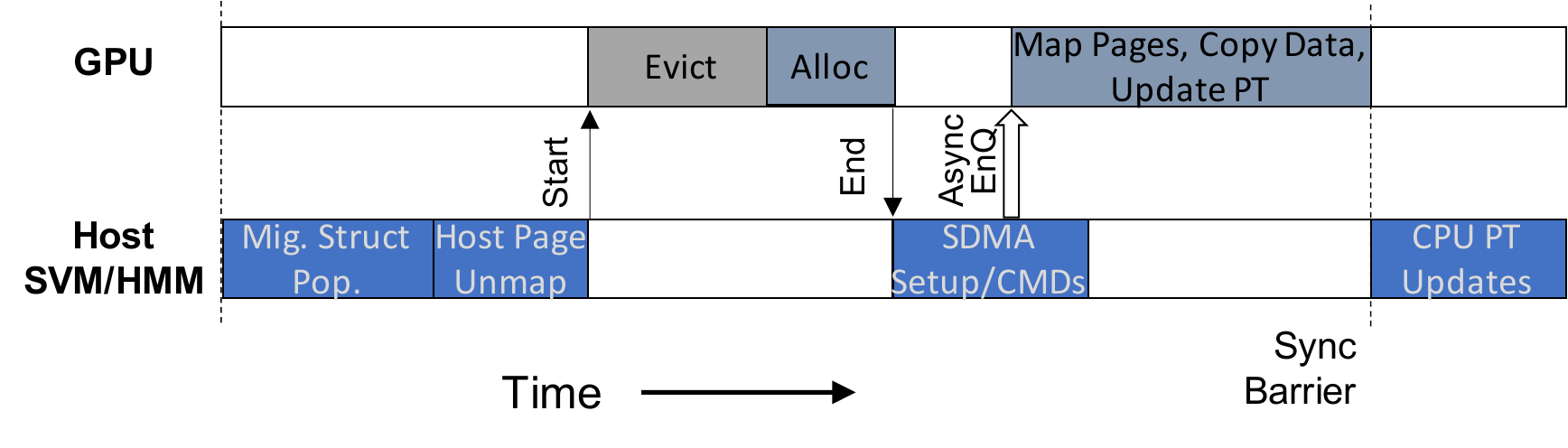}
\vspace{-20pt}
\caption{Timeline of range migration for a serviceable fault. ``Evict" only occurs if there is insufficient space for ``Alloc".}
\vspace{-10pt}
\label{fig:migrate_timeline}
\Description{Timeline of migration operations for a serviceable fault.}
\end{figure}

Each serviceable fault triggers SVM to migrate the range in which the faulting page is.  SVM keeps track of the residency of each range and uses it to determine the migration direction (e.g., host-to-device, device-to-host, or device-to-device). Due to page limits, we focus on the host-to-device migration as it is the most important in GPU computing. 

To migrate a range from host to device, the SVM driver provides HMM the start and end addresses of the range and obtains a list of source physical frame numbers (PFNs) necessary for data migration. HMM further leverages the built-in memory management in the Linux kernel for host memory management and  page table operations.

Figure~\ref{fig:migrate_timeline} shows the timeline of the host-to-device migration visible in the SVM driver. Essentially, the SVM  driver sends commands to the host and the device and synchronizes their operations. The commands to the host/HMM include obtaining source PFNs, and performing page unmapping and page table updates, and commands to the device include allocating ranges and pages, initiating direct memory copy, and performing page mapping and page table updates.  Note that the System Direct Memory Access (SDMA) copy is asynchronous and used for page content copy, GPU page mapping and unmapping, and page table updates. The associated cost partly overlaps with SDMA setup and command issuing on the SVM driver. 

During the GPU memory allocation, if there isn't enough available space for the range to be migrated, SVM must first evict one or more residing ranges, as shown in Figure~\ref{fig:migrate_timeline}.  SVM employs the least recently faulted (LRF) policy to determine the next victim range and continues to evict ranges until the available space becomes sufficient. It's important to note that eviction is costly, involving various operations such as page mapping, unmapping, and content copying as in migration, albeit in the opposite direction. We show the quantitative costs in detail in section~\ref{sec:costs}.  

\subsection{Overall SVM Architecture}

\begin{figure}
\centering
\includegraphics[width=\linewidth]{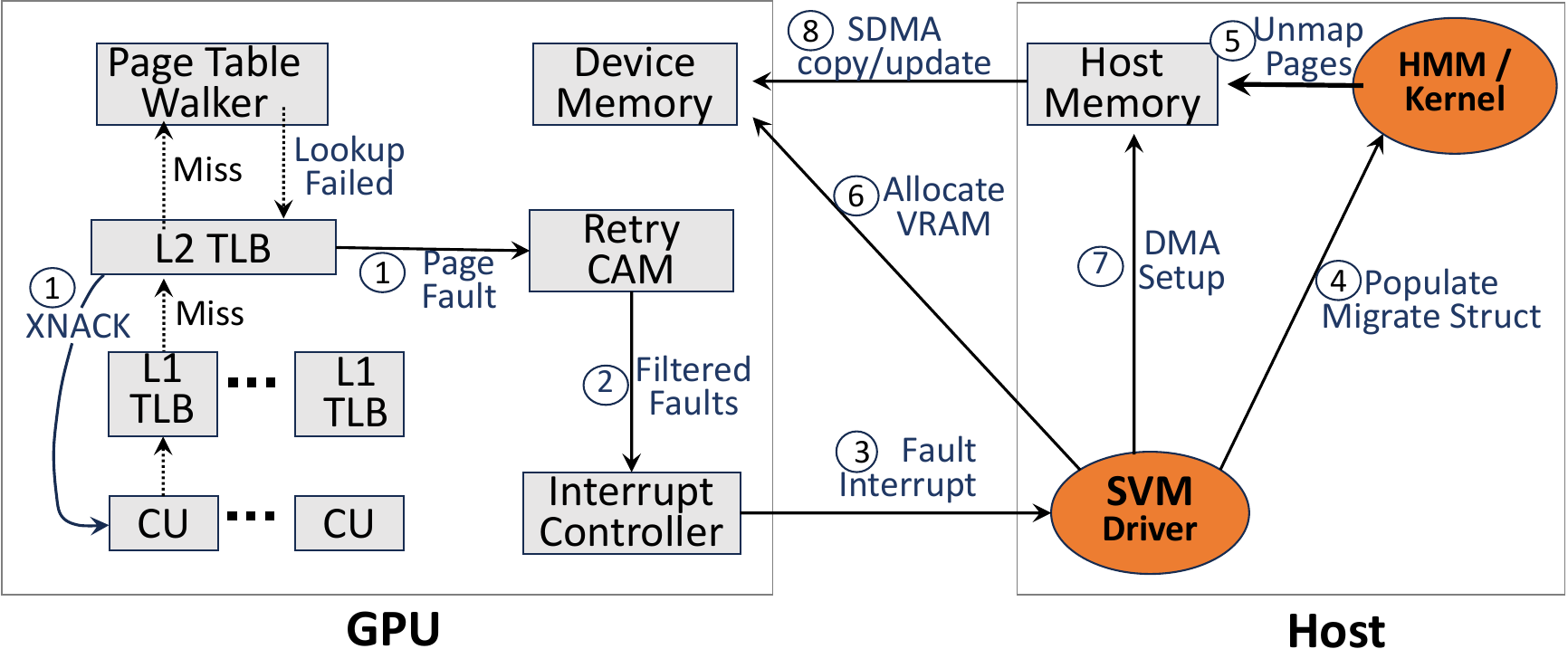}
\caption{SVM Architecture. The steps are components' interactions in response to a page fault originating from a compute unit (CUs).}
\vspace{-15pt}
\label{fig:svmarch}
\Description{The path of a serviceable fault originating from GPU hardware through SVM handling.}
\end{figure}

Figure~\ref{fig:svmarch} presents the SVM architecture, and how its modules interact to service a serviceable page fault originating from a compute unit (CU). A page fault occurs if address translation using TLBs and page tables fails. Upon the fault, the L2 TLB sends a ``translation negative acknowledgment'' (XNACK) back to the CU and writes an interrupt cookie to an on-device buffer Content-Addressable-Memory (CAM) \textcircled{\raisebox{-1.2pt}{1}}, in which faults on the same pages can be filtered. The interrupt controller reads from the buffer \textcircled{\raisebox{-1.2pt}{2}} and passes these cookies along to the SVM driver for servicing \textcircled{\raisebox{-1.2pt}{3}}. Meanwhile, the CU retries the faulted access until translation succeeds.

\begin{figure*}
\centering
\subfloat[][STREAM]{\includegraphics[width=.3\linewidth]{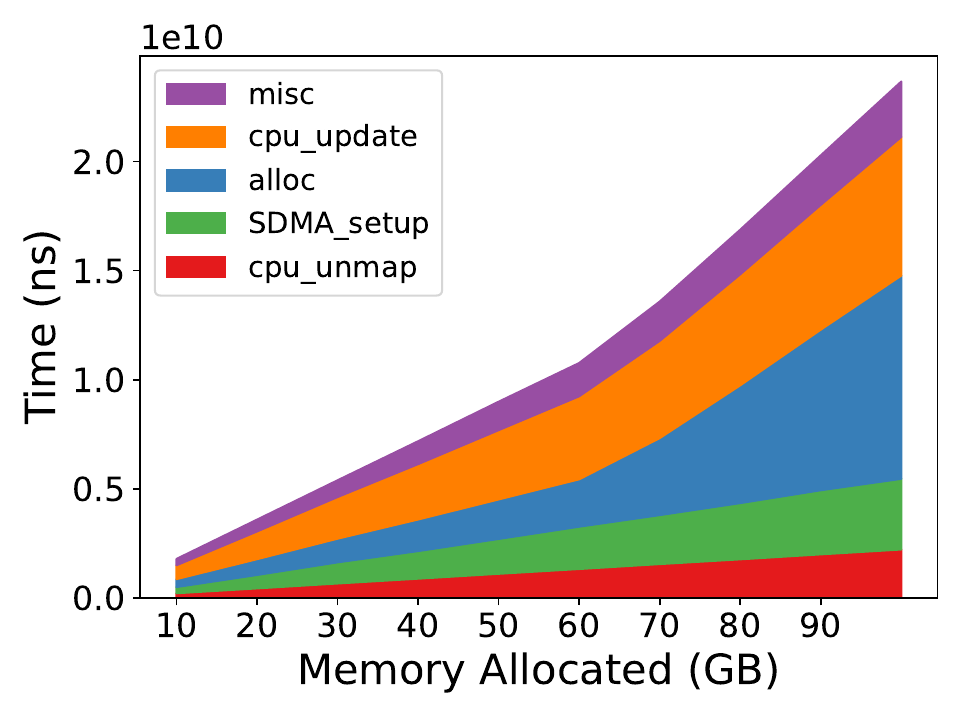} }
\subfloat[][Jacobi2d]{\includegraphics[width=.3\linewidth]{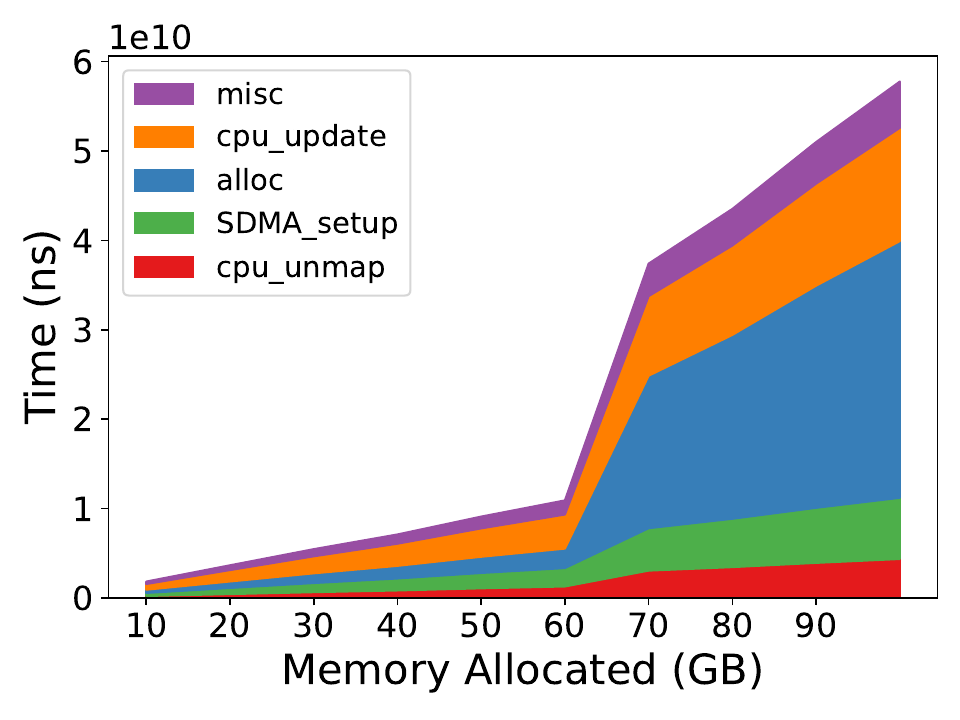} }
\subfloat[][SGEMM]{\includegraphics[width=.3\linewidth]{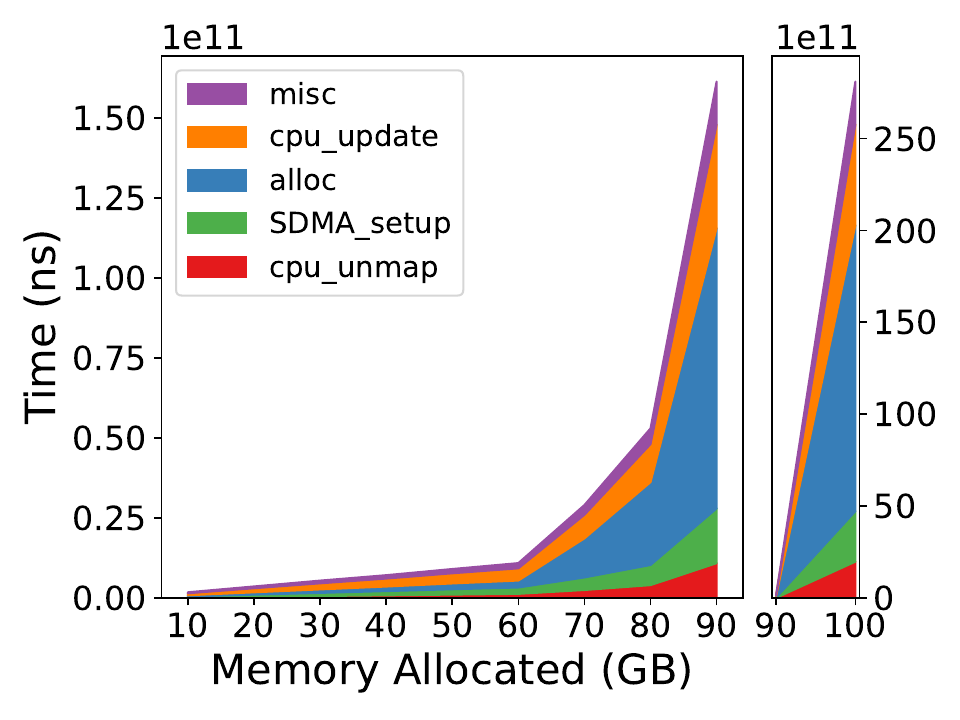} }
\vspace{-5pt}
\caption{The cost of SVM UM management and range migration. SGEMM is shown in two windows as the magnitude of the second visually erases the first.
}
\vspace{-10pt}
\label{tioga-driver-breakdown}
\Description{The cost of key SVM operations for select applications across problem sizes.}
\end{figure*}

The SVM driver running on the host decodes the interrupt cookie. This cookie contains the faulting page address, the timestamp, and the access type. The driver determines if the fault is serviceable, i.e., neither timed out nor a duplicate range.

For a serviceable fault, SVM creates a \texttt{migrate\_vma} structure and passes it to HMM along with the corresponding range in order to get the host source page PFNs populated \textcircled{\raisebox{-1pt}{4}}. HMM performs a page table walk over the range and unmaps all pages on the host \textcircled{\raisebox{-1pt}{5}}. The SVM driver then allocates memory on the GPU device as needed for the migration \textcircled{\raisebox{-1.2pt}{6}}. Next, the driver sets up the System Direct Memory Access (SDMA) mapping and issues commands \textcircled{\raisebox{-1.4pt}{7}} to asynchronously copy data, perform paging, and update GPU page tables \textcircled{\raisebox{-1.2pt}{8}}. Migration concludes at the synchronization point. 

SVM architecture shows distinct strategies for fault handling and unified memory management compared to NVIDIA's Unified Virtual Memory (UVM), which has garnered relatively more research attention~\cite{allen-ipdps21, allen-sc21, kim-asplos20}. Table~\ref{tb:SVM-UVM} presents the key differences. Unlike UVM which batches page faults in a buffer and  handles the batch when the buffer is full, SVM receives a single fault each time and handles it immediately. Such a strategy has two main advantages. First, serviceable faults are serviced immediately to reduce the turnaround time for individual accesses. Second, duplicate page faults are quickly identified and dismissed from servicing. The main disadvantage is that the SVM driver is heavily loaded with fault interrupts, even after some faults are filtered in the CAM buffer on the GPU side. 

SVM manages the UM at the range granularity for allocation, deallocation, migration, and eviction. While being varying sizes, a range is typically orders of magnitude larger than a VABlock (2 MB) used in UVM.  If all migrated data are to be used, migrating an entire range effectively amortizes data access latency and fully utilizes the host-device interconnect bandwidth. The range granularity is beneficial for scenarios when GPU memory is not oversubscribed or for applications whose data are not evicted before use. Otherwise, it causes significant performance issues  for two main reasons. First, GPU memory gets exhausted faster with data not immediately needed, requiring evictions to make space for subsequent migrations. Second, it causes severe thrashing, i.e., data migrated but evicted before use must be migrated again. Thrashing degrades performance in two ways: wasting time in migrating and then evicting unused data, and more importantly increasing the migration frequency. Case studies of severe thrashing are presented in Section~\ref{sec:thrashingcost}. 

\begin{table}
    \begin{centering}
    \begin{tabular}{|l|l|l|}
        \hline
        \textbf{UM Feature} & \textbf{ SVM} & \textbf{ UVM} \\ \hline
        \textbf{Fault batching} & No & Yes  \\ \hline
        \textbf{Fault handling} & Single fault & Fault batch$^a$  \\ \hline
        \textbf{UM (De)alloc} & Range ($\in$[4KB, 1GB]) & VABlock (2MB)\\ \hline
        \textbf{Migration unit} & Range & Page$^b$ \\ \hline
        \textbf{Eviction unit} & Range & VABlock \\ \hline
        \textbf{Eviction Policy} & \multicolumn{2}{|c|}{Least Recently Faulted} \\ \hline
    \end{tabular}

    \vspace{2pt}
    \noindent\footnotesize{$^a$ A batch consists up to a system-configurable 256 faults.}\\
    \noindent\footnotesize{$^b$ 64 KB without prefetching, and up to a VABlock  with prefetching.}\\
    \end{centering}
    \caption{ SVM vs.  UVM.}
    \vspace{-28pt}
    \label{tb:SVM-UVM}
\end{table}

\subsection{SVM UM Management Costs}\label{sec:costs}

We quantitatively analyze the  cost for SVM UM management. We use Systemtap~\cite{systemtap} to dynamically instrument and trace the  SVM driver functions and events. We run each instance twice for data collection: first to measure the timing of SVM driver functions, and then to capture events such as faults, migrations, and evictions. 

Here we focus on the major cost items during fault servicing and corresponding migration, based on Figure~\ref{fig:migrate_timeline}, and ignore others including fault receiving, preprocessing, and filtering as their costs are relatively small and negligible.
\begin{itemize}
    \item \textbf{cpu\_unmap}: collect and unmap host pages.
    \item \textbf{SDMA\_setup}: create SDMA mappings, and issue SDMA commands to perform copy, mapping, and page updates.
    \item \textbf{alloc}: allocate physical VRAM on the device. Note that this item includes the cost of eviction if there is insufficient space for allocation.
    \item \textbf{cpu\_update}: update CPU page table with new mappings if migration succeeded or restore old mappings if failed.
    \item \textbf{misc}: migrate page meta-data, non-overlapped SDMA copy, and free copy mappings.
\end{itemize}
    
\textbf{cpu\_unmap} and \textbf{cpu\_update}  manage pages and page tables on the host side, and the actual data movement is encapsulated in  \textbf{SDMA\_setup} and \textbf{misc}. All these costs are visible on the host and are reflected in the application's execution time. 

\begin{table*}
\centering
\begin{tabular}{|l|l|l|l|}
     \hline
     \textbf{Benchmarks} & \textbf{Description} & \textbf{Domain} & \textbf{Source} \\ \hline
     STREAM & Triad-only. Scaled dot product of two vectors. & Synthetic & RAJAPerf~\cite{rajaPerf} \\ \hline
     Conv2d & Full convolution in a 2D space with varying weights. & Machine Learning & RAJAPerf~\cite{rajaPerf} \\ \hline
     Jacobi2d & Forward then backwards adjacent convolution with equal weights. & Machine Learning & RAJAPerf~\cite{rajaPerf} \\ \hline
     BFS & Breadth First Search graph traversal from randomly selected node. & Graph Traversal & EMOGI~\cite{emogi} \\ \hline
     SYR2K & Symmetric rank-2k update from ROCBLAS & Linear Algebra & rocBLAS~\cite{rocblas} \\ \hline
     SGEMM & General matrix-matrix product from ROCBLAS & Linear Algebra & rocBLAS~\cite{rocblas} \\ \hline
     MVT & Matrix-vector product followed by matrix-transpose-vector product. & Linear Algebra & RAJAPerf~\cite{rajaPerf} \\ \hline
     GESUMMV & Sum of two scaled matrix-vector products. & Linear Algebra & RAJAPerf~\cite{rajaPerf} \\ \hline
\end{tabular}
\caption{Diverse benchmarks from multiple domains.}
\vspace{-20pt}
\label{tb:apps}
\end{table*}

Figure \ref{tioga-driver-breakdown} shows the cost items for three representative applications over various problem sizes ranging from small to large enough to oversubscribe GPU memory by 56\%. Note that each cost item is accumulated over all the migrations in the kernel. From these figures, we make some key observations. 

\begin{itemize}
    \item The total cost increases with problem size as expected because the number of migrations increases. However, the applications exhibit distinct growth trends. STREAM displays two linear segments separated by oversubscription, with the slope of the second segment being slightly larger. Both Jacobi2D and SGEMM present three or more segments. For Jacobi2D,  the second segment's slope is the largest where GPU memory is oversubscribed by less than 10\%. For SGEMM, the last segment's slope is significantly larger, surpassing the others by orders of magnitude. 
    \item For small problem sizes without oversubscribing GPU memory, \textbf{cpu\_update} is the largest individual component, followed by \textbf{SDMA\_setup} and \textbf{alloc}. These three account for roughly 76\% of the overall cost for all three applications. 
    \item While all cost items increase under oversubscription, \textbf{alloc} increases the most and becomes dominant across the applications. This increase is due to the evictions for freeing GPU memory. Eviction comprises all other items in the opposite direction, thus is costly. The slightly larger slope of the second segment in STREAM suggests only a small  number of evictions, while the drastically larger slopes in Jacobi2D and Sgemm suggest higher numbers of evictions. 
\end{itemize}

These results indicate significant overhead for UM and demand paging. When GPU memory is not oversubscribed, the actual data movement across the host and device memory domains only accounts for less than half of the overall cost, and is smaller than the sum of UM management items including host and device mapping and unmapping and page table updates. More seriously, the UM management overhead  increases substantially once GPU memory is oversubscribed, and becomes extremely high for applications such as SGEMM. We examine how the overhead impacts the performance of various applications in the following section.
\section{Workload Performance and Profiles}
\label{sec:workloads}

We examine diverse GPU workloads and study how their performances vary with GPU memory oversubscription. We further inspect their migration and eviction profiles, and use them to explain the performance change. The applications represent multiple domains as listed in Table \ref{tb:apps}. Some are directly from AMD ROCm implementations (e.g., rocBLAS SYR2K and SGEMM), and others are ported from RAJAPerf~\cite{rajaPerf} implementations using Heterogeneous-computing Interface for Portability (HIP) APIs. We modify responding allocations to utilize managed memory. We have examined over a dozen applications but have only included those with complete data across different problem sizes.

\subsection{Performance Impacts of Oversubscription}

We use the term Degree of Oversubscription (DOS) to quantify how much  memory is used beyond the GPU's available capacity for unified memory (UM). DOS is defined as ${used\_size}/{available\_size} \times 100$. Based on this definition, a DOS value exceeding 100 indicates GPU memory oversubscription.

\begin{figure}
\centering
\includegraphics[width=0.85\linewidth]{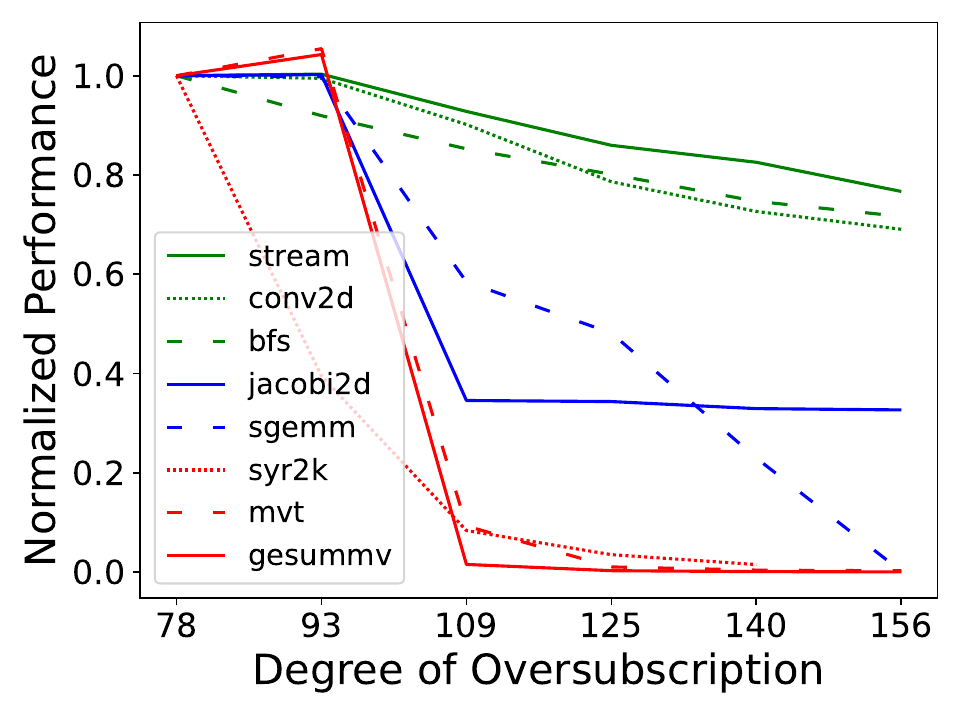}
\vspace{-10pt}
\caption{Performance decreases with the degree of oversubscription for various applications under SVM.}
\vspace{-10pt}
\label{fig:app_perf}
\Description{Performance trends of all applications as they oversubscribe GPU memory.}
\end{figure}

Figure \ref{fig:app_perf} shows how performance varies with DOS  with demand migration across the  applications. Application performance is measured using throughput, which can be compute rate (FLOPs per Second) or memory throughput (GBs per second), and is normalized to that at $\texttt{DOS}=78$. We use normalization here to emphasize the change in an application's performance relative to the problem size. While all applications' performances decrease monotonically as DOS increases, they exhibit different patterns. We group the patterns into three categories.

\textbf{Category \uppercase\expandafter{\romannumeral1\relax}}: Performance declines moderately as DOS increases. STREAM and Conv2d  belong to this category, though their rates of decrease are different. When oversubscribing GPU memory, applications in this category experience the least impact with demand migration. 

\begin{figure*}
\centering
\subfloat[][STREAM]{\includegraphics[width=.24\linewidth]{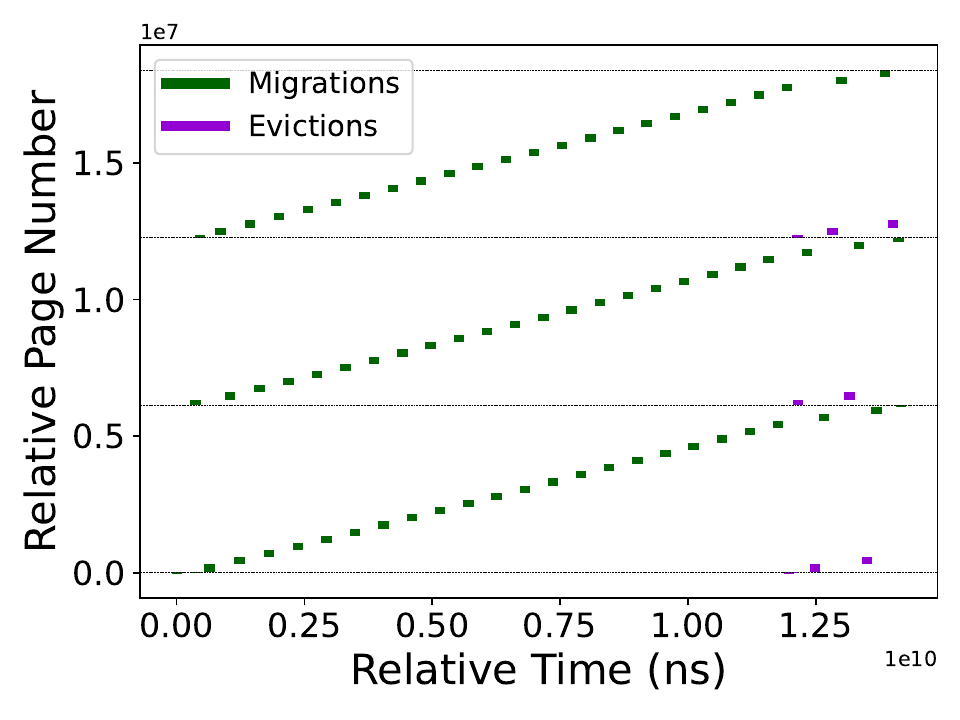} }
\subfloat[][Conv2d]{\includegraphics[width=.24\linewidth]{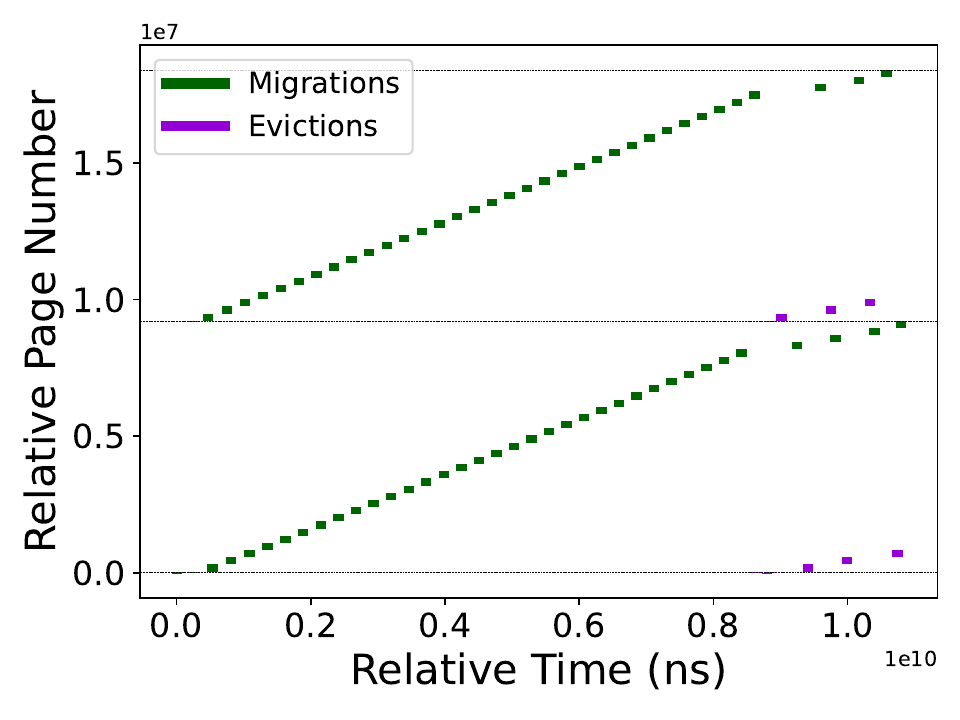} }
\subfloat[][BFS]{\includegraphics[width=.24\linewidth]{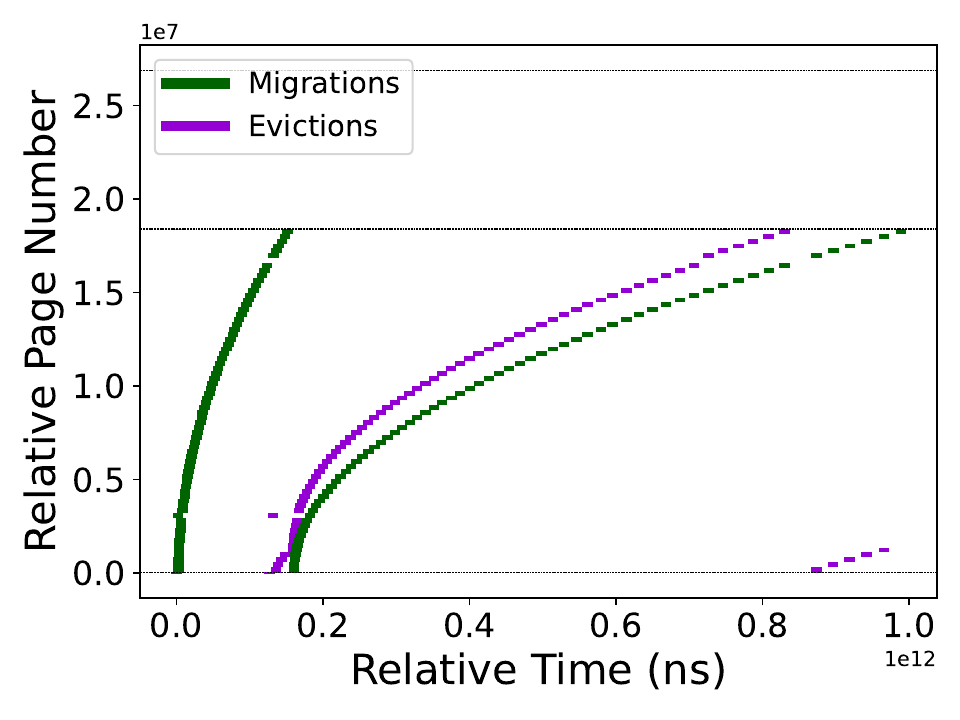} }
\subfloat[][Jacobi2d]{\label{subfig:j2d-ap}\includegraphics[width=.24\linewidth]{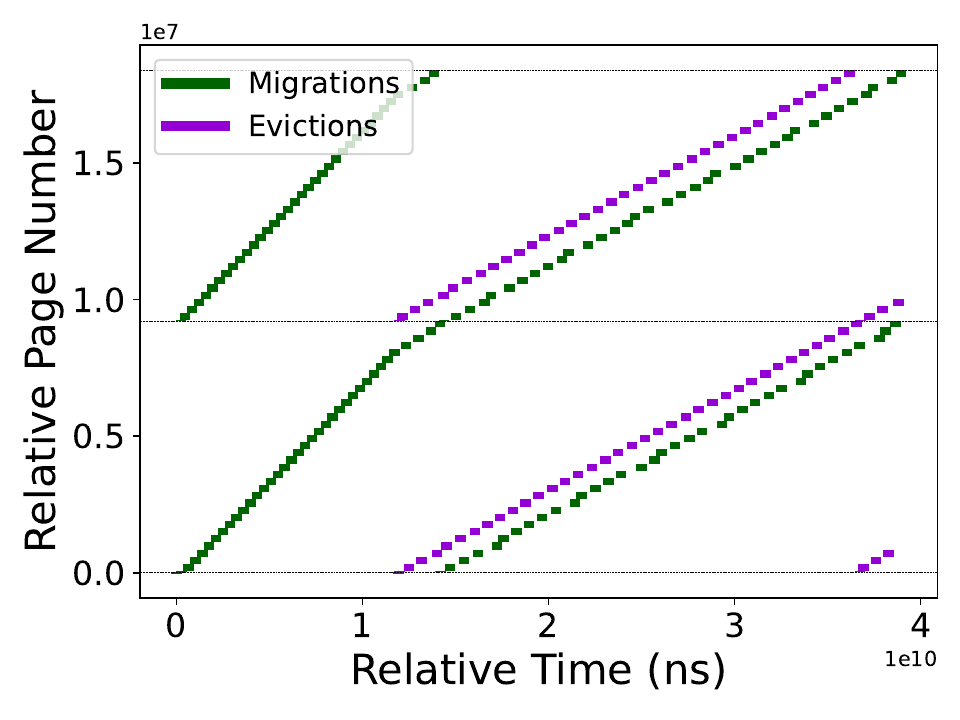} }
\\
\subfloat[][SGEMM]{\label{subfig:sgemm-ap}\includegraphics[width=.24\linewidth]{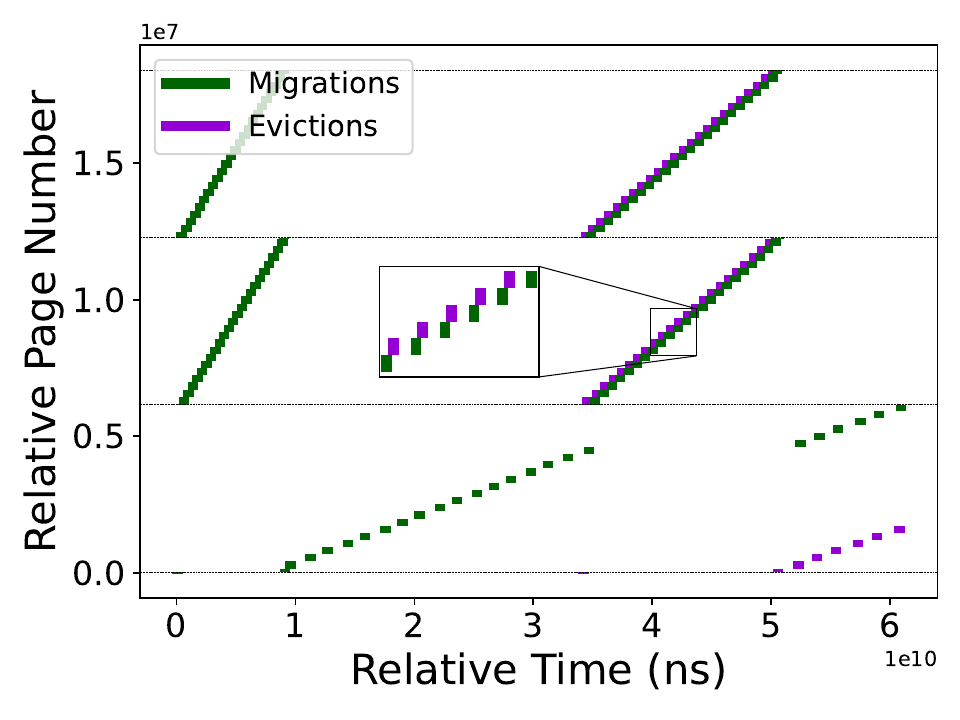} }
\subfloat[][SYR2K]{\includegraphics[width=.24\linewidth]{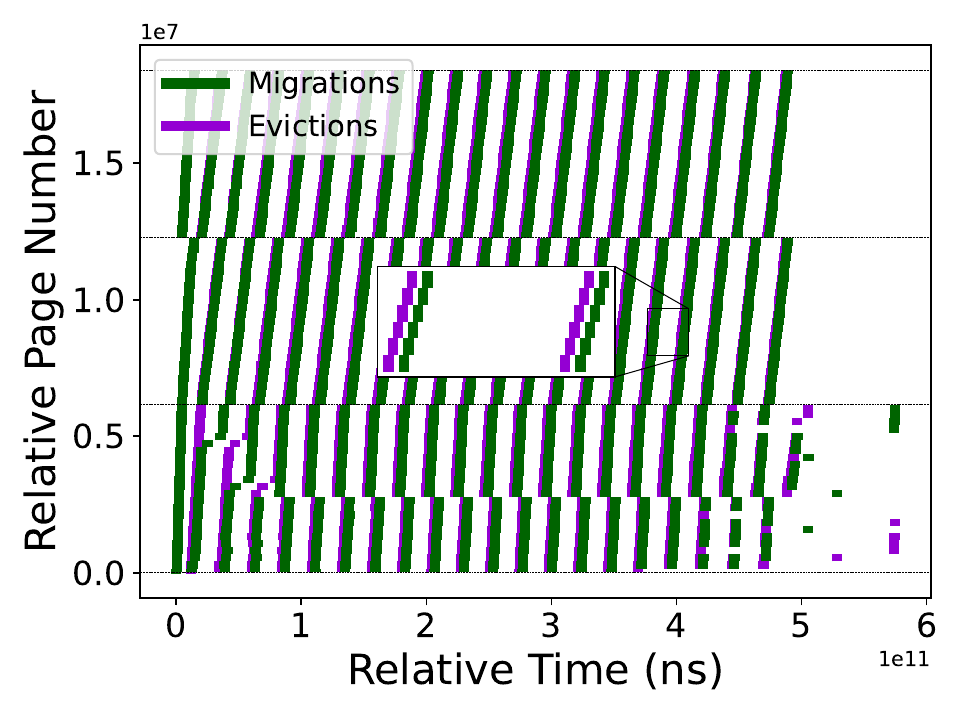} }
\subfloat[][MVT]{\includegraphics[width=.24\linewidth]{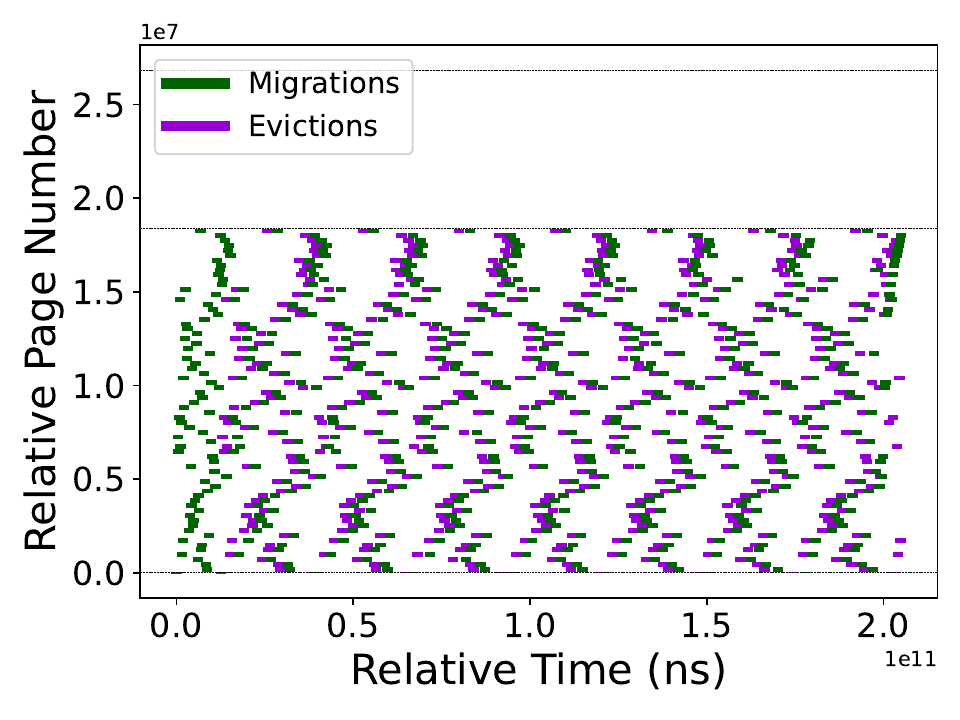} }
\subfloat[][GESUMMV]{\includegraphics[width=.24\linewidth]{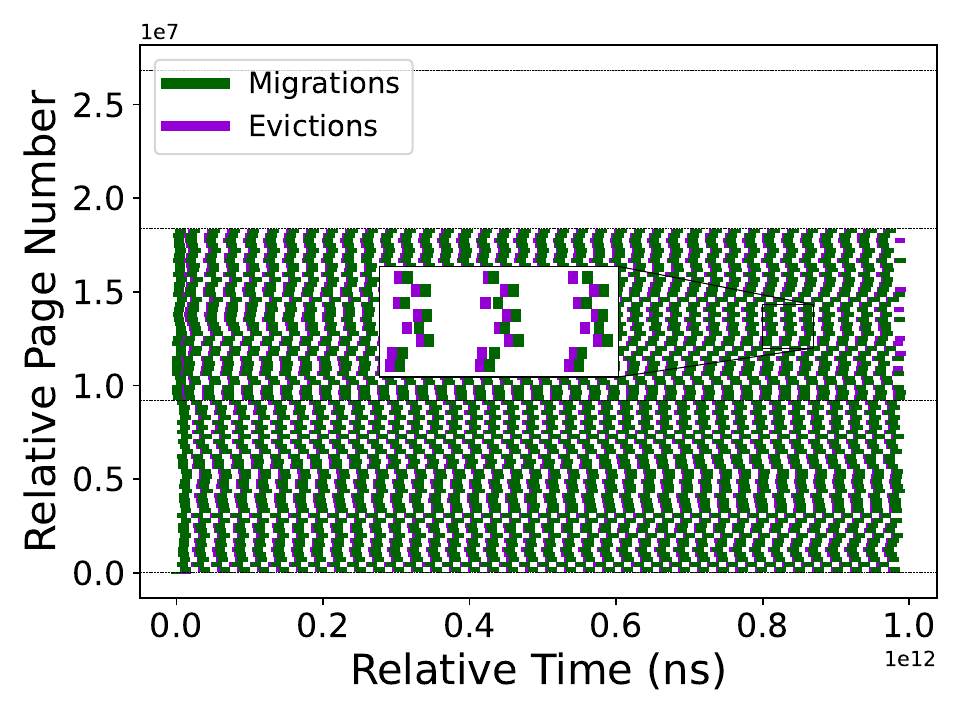} }
\vspace{-5pt}
\caption{Migrations and evictions over execution time at $DOS=109$. The y-axis is subdivided by allocation boundaries. White spaces are respective to smaller allocations. Though invisible, they also experience migration and evictions. Their invisibility is a result of the presentation: enlarging these data points causes the currently visible ones to occupy a solid-filled space. 
}
\vspace{-10pt}
\label{fig:access-patterns}
\Description{Migration \& eviction timelines for each application.}
\end{figure*}

BFS appears to be in category \uppercase\expandafter{\romannumeral1\relax}, warranting explanation. BFS's execution depends on the input graph and the start node. Our case study uses a randomly generated graph with 10\% of possible edges and a randomly selected start node.  While accesses to nodes and edges are expected to be random, this randomness is confined within  ranges, and  the accesses across the  ranges still follow a linear pattern. 

\textbf{Category \uppercase\expandafter{\romannumeral2\relax}}: Performance declines  significantly once DOS surpasses 100, and then minimally changes thereafter. Jacobi2D belongs to this category and its performance decreases to about 40\% at $DOS=109$.  

\textbf{Category \uppercase\expandafter{\romannumeral3\relax}}: Performance drops close to zero when DOS surpasses 100 or more. The decline can be abrupt as seen in GESUMMV and MVT, or gradual as in SGEMM. When oversubscribing, applications in this category experience the most impact using the SVM on-demand migration.

What are the factors responsible for the varying performance differences among the applications as DOS changes? To answer this question, we  inspect the migration and eviction profiles resulting from the applications' interaction with SVM. Limited by the long time needed to gather experimental results, Figure~\ref{fig:app_perf} only displays performance data for DOS values up to 156. Other questions that naturally arise are: What are the performance of applications like STREAM and Jacobi2D as DOS continues to increase? We establish the answers using insights from fine-grain application profiles presented next. 

\subsection{Migration and Eviction Profiles}
The performance impact is determined by the  complex interplay between the application's memory request and SVM UM management. As noted before,  performance degradation under oversubscription is primarily attributed to eviction and, even more significantly, to thrashing. Eviction is on the critical path, meaning eviction is only initiated by the migration request in the opposite direction, which is blocked until eviction is completed. Eviction doubles the cost of migration and   delays the migration and computation. In general, the more an application evicts, the larger the performance loss it suffers. 

In an on-demand migration memory model, eviction is inevitable for problem sizes that exceed the GPU's physical memory. While eviction directly results in performance loss, some evictions are more costly.  Eviction has two types: permanent eviction, which displaces data no longer needed, and premature eviction, which displaces data required for current or future computation. Permanent evictions simply increase migration costs, while premature evictions further lead to thrashing, which has a compounding effect by increasing the eviction-to-migration ratio and the number of migrations.

Premature evictions occur in applications with certain temporal and spatial access patterns. The temporal pattern involves data reuse, and particularly the repeated traversal of one or more memory allocations. Such a pattern is commonly found in algorithms with nested loops, such as BLAS-2 and BLAS-3 algorithms.  Any eviction of the repeatedly traversed allocations is premature, necessitating the subsequent migration. The spatial pattern involves successive accesses of a small amount of data that is distributed across the ranges of the same allocations. Applications displaying such patterns rapidly fill the GPU's memory, leading to frequent evictions,  of which  a significant portion is premature.

Figure \ref{fig:access-patterns} shows migration and eviction profiles at $DOS=109$ across the applications. The data are collected using Systemtap as described in Subsection \ref{sec:costs}. It is worth noting that the profiles only reveal partial information, i.e., transfers of ranges across host-device interconnect that involve the SVM driver. They are unable to show access to data within the ranges or data re(use) on the GPU device. Missing information such as the number of faults a migration satisfies is crucial for identifying performance bottlenecks and opportunities for optimization. We discuss it in detail in Subsection~\ref{sec:faultlevel}. 

Applications in Category \uppercase\expandafter{\romannumeral1\relax} such as STREAM and Conv2d involve only permanent evictions. The ranges of each allocation are migrated in a linear streaming fashion, and all allocations are concurrently accessed. Once GPU memory is oversubscribed, ranges migrated the earliest are evicted successively. These applications don't have data reuse. 

BFS iterates over multiple GPU kernels using the same data and thus incurs premature evictions. Determined by the linear traversal of the edge list's ranges and the minimal computation, BFS's degradation complies more with category \uppercase\expandafter{\romannumeral1\relax}.

Applications in Category II such as Jacobi2D also exhibits linearly progressed range migrations over all of its  allocations. However, Jacobi2D experiences premature evictions: ranges are evicted and then re-migrated a short time later. Jacobi2d iterates the same data accesses and computations, and Figure~\ref{fig:access-patterns}d illustrates two iterations. Execution with a larger problem size  should have the same migration and eviction profiles but with an earlier onset of eviction in the initial iteration. 

There are two subtypes of applications in Category III. One type includes applications SGEMM and SYR2K that exhibit linearly progressed range migrations and premature evictions for allocations, similar to Jacobi2d. A key feature is that their prematurely evicted data are intensively reused for computation at present and in the future, manifested by immediately re-migrating the same ranges after evicting them.

The other subtype in Category III such as MVT and GESUMMV display  spatial patterns where successive data accesses are dispersed across the allocations as in matrix transpose.  A  large number of unique ranges are migrated and evicted simultaneously. GPU memory is quickly filled and evictions occur very early in the application's execution. Although  MVT and GESUMMV do not reuse data as intensively as  Sgemm and SYR2K , they experience similar thrashing. GESUMMV suffers more thrashing than MVT with two large allocations instead of one. 

What is the performance of applications like STREAM and Jacobi2D as DOS continues to increase? We derive that STREAM's performance asymptotically approaches half of the highest performance when $DOS<100$. As DOS increases, every migration, except the ones before the GPU memory is fully occupied, requires an eviction, which involves the same operations in the opposite direction. When the eviction-to-migration ratio approaches 1, performance halves. Using the same analysis, we derive that Jacobi2D's performance approaches 0.36. 
\begin{figure}
    \centering
    \includegraphics[width=0.80\linewidth]{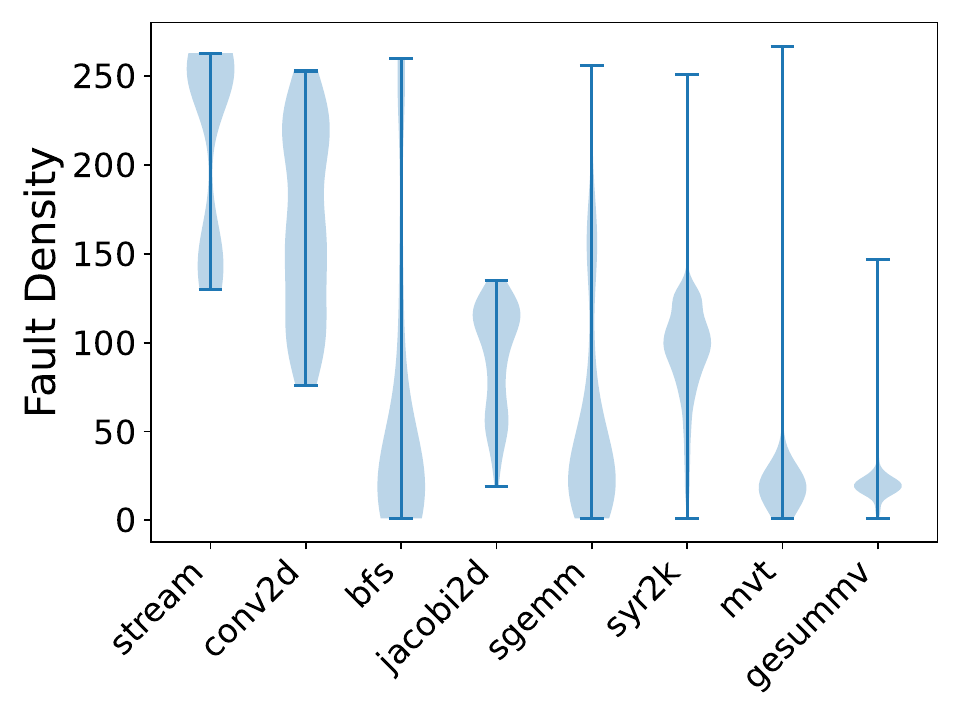}
    \vspace{-10pt}
    \caption{Overall fault densities for application executions with problem sizes at $DOS=109$.}
    \vspace{-12pt}
    \label{fig:fault-density-violin}
    \Description{Distribution of fault densities for each application.}
\end{figure}

\begin{figure*}
\centering
\subfloat[][STREAM]{\includegraphics[width=.33\linewidth]{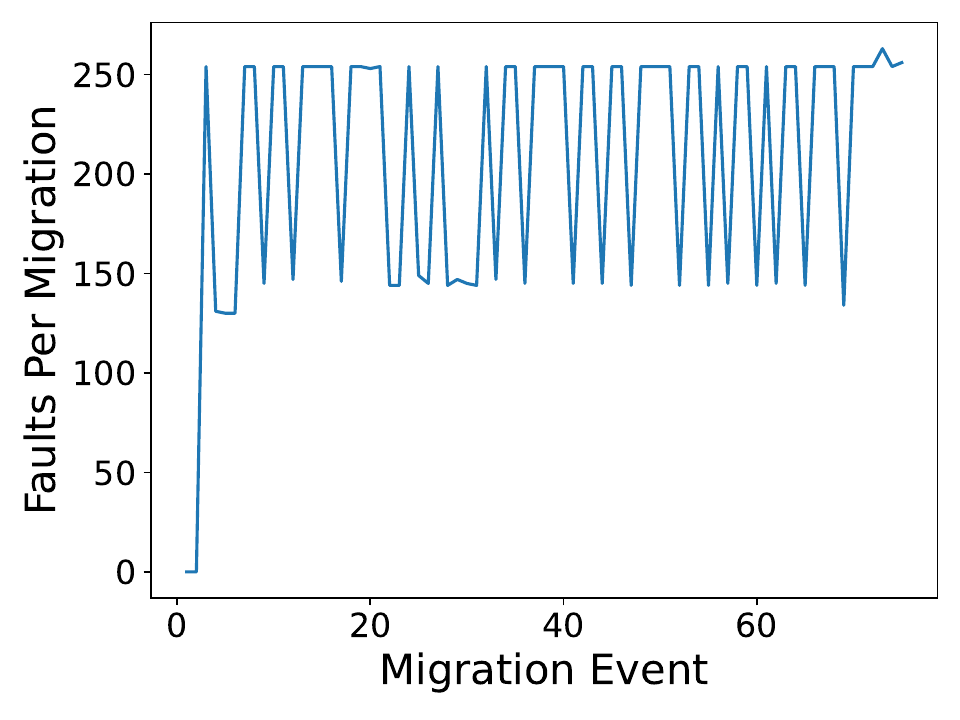}}
\subfloat[][SGEMM]{\includegraphics[width=.33\linewidth]{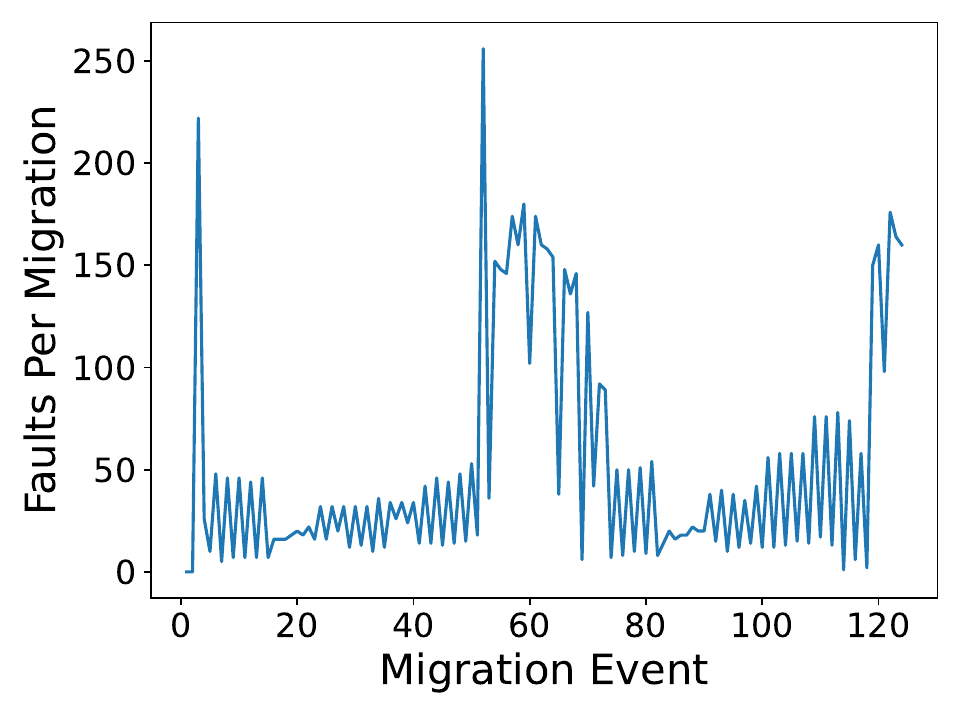} }
\subfloat[][GESUMMV]{\includegraphics[width=.33\linewidth]{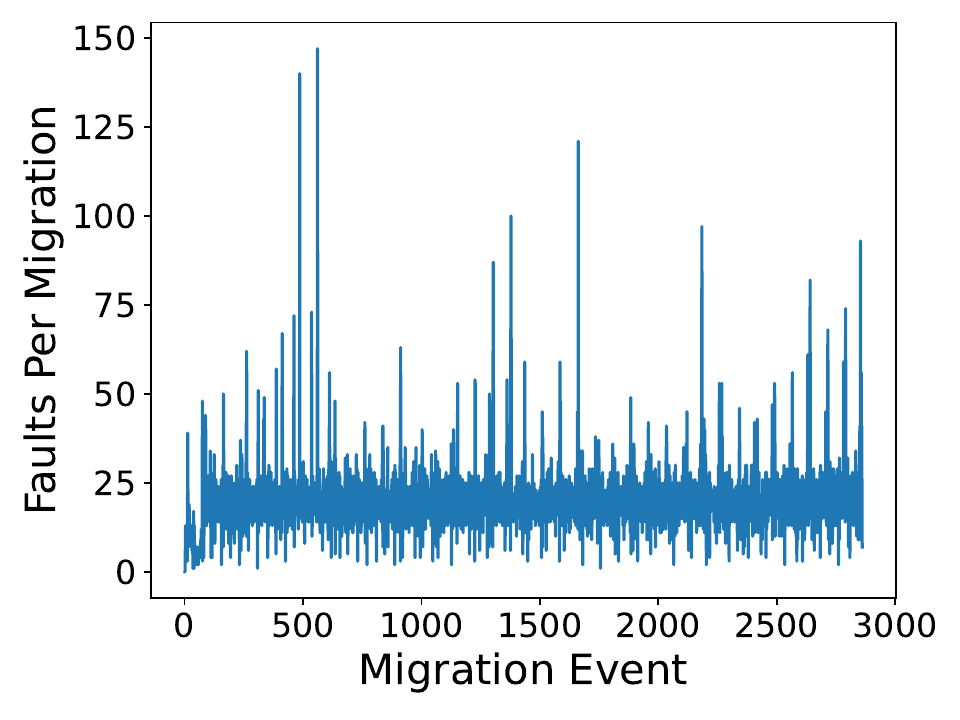} }
\\
\vspace{-7pt}
\subfloat[][STREAM]{\includegraphics[width=.33\linewidth]{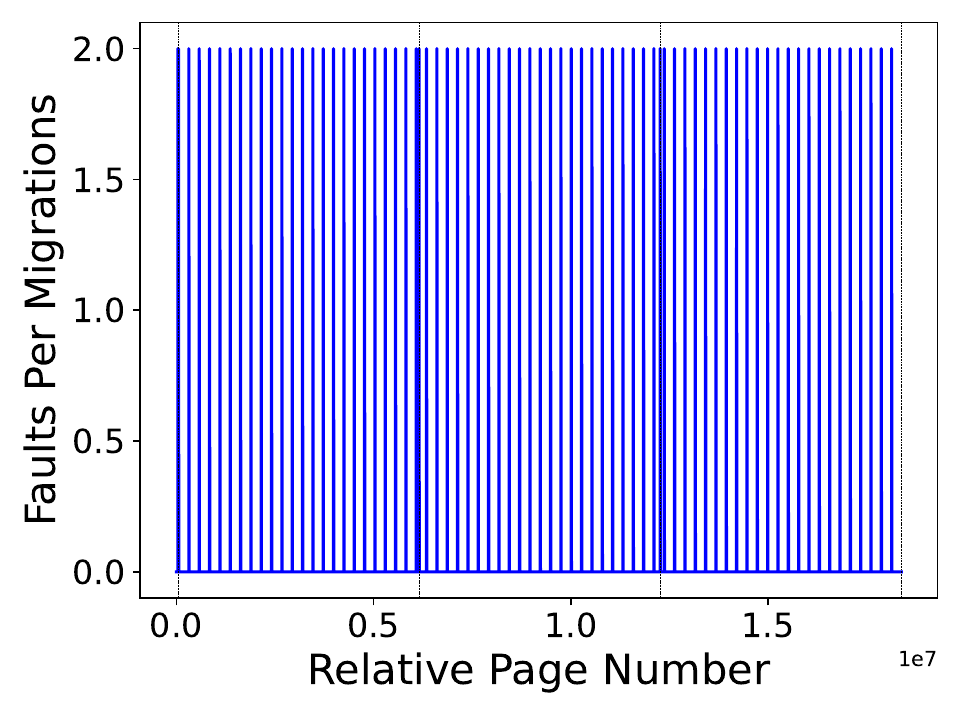}}
\subfloat[][SGEMM]{\includegraphics[width=.33\linewidth]{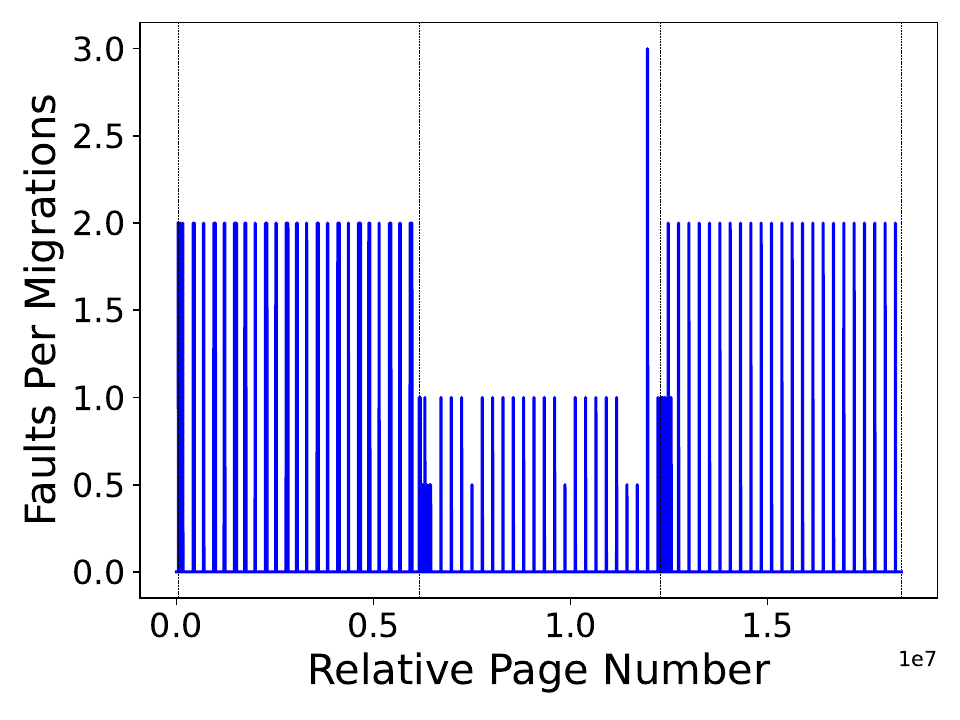}}
\subfloat[][GESUMMV]{\includegraphics[width=.33\linewidth]{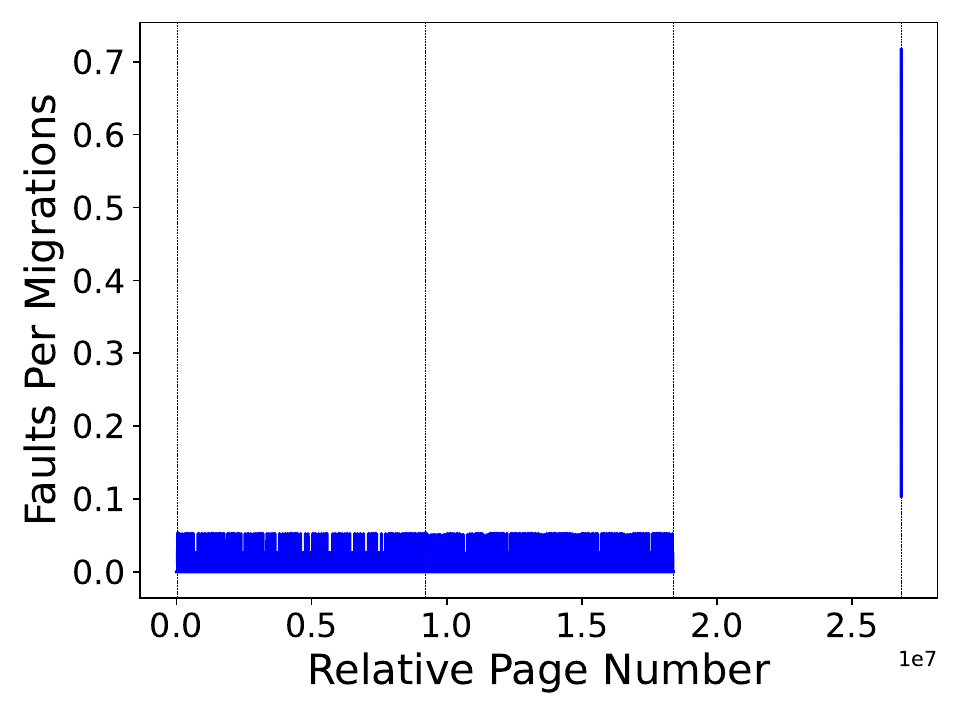}}
\vspace{-5pt}
\caption{Variation of fault density over time (a-c) and over allocations (d-f) at $DOS=109$.
}
\vspace{-12pt}
\label{fig:fault-density-plots}
\Description{Individual fault densities for all migrations (a-c) and fault densities for all pages (d-f) for select applications.}
\end{figure*}

\subsection{Fine-Grain Fault Behaviors}\label{sec:faultlevel}
\label{sec:internal}
Migration and eviction profiles offer incomplete information and cannot entirely quantify the performance differences among the applications.  Here, we delve into the detailed fault behaviors and assess how effectively they are handled by migrations.

We use \textbf{\textit{fault density}} to refer to the number of faults that are satisfied by a given migration. Here faults include both serviceable and dismissed to reflect the data request of applications. The higher the fault density is, the more effective a migration is. There are two conditions for an application to obtain high fault density. First, the application must consecutively request a large amount of data in the same range. Second, these accesses must occur simultaneously or in a small enough time frame, e.g., the time taken to service a fault. Applications with linearly progressed access inherently meet the first condition, but other access patterns may also meet it. The second condition is predominately driven by the arithmetic intensity, typically measured by the compute per data access~\cite{williams2009roofline}, of an application. Greater arithmetic intensity results in lower fault density by enlarging the time window between accesses.

Figure \ref{fig:fault-density-violin} presents the fault densities across the applications. Overall, applications in Category I such as STREAM and Conv2d have the highest fault densities. They both have linearly progressed accesses. Between them, Conv2D has a somewhat lower fault density with its higher arithmetic intensity. Next comes Jacobi2d in Category II. While Jacobi2d also has linearly progressed accesses, it involves evictions which enlarges the time frame for the same number of faults. Applications in Category III such as MVT and Gesummv have the lowest fault densities for their successive accesses are distributed over the ranges. BFS is an exception, with linearly progressed accesses and low arithmetic intensity but a very low average fault density as those in Category III. This is explained by its random and sparse accesses of the edges and nodes within the ranges.  

Each application's fault density has a certain range and distribution. Figure~\ref{fig:fault-density-plots} shows how the fault density varies over time (Figure~\ref{fig:fault-density-plots}a-c) and  over allocations using three applications (Figure~\ref{fig:fault-density-plots}d-f). STREAM's fault density largely falls in $[150, 250]$ over time. SGEMM has a lower average fault density below 50 for it is computationally intensive. The spikes correspond to periods during which data migration occurs without concurrent computation. GESUMMV's fault density varies over time and fluctuates around 20 because it successively accesses to data distributed over ranges. 
 
As shown in Figure~\ref{fig:fault-density-plots}d-f, only specific page numbers have a non-zero fault density and trigger migrations, while  others encounter zero faults because their requests are already satisfied by the range migrations. Migration-triggering pages in STREAM and SGEMM are uniformly distributed across their allocations, complying with their linearly progressed data accesses.  These pages correspond to those located at the start of the ranges. Note that each bar encapsulates a large number of such pages.  The average faults per migration is 2, indicating duplicate faults. Migration-triggering pages in GESUMMV are not only  more densely distributed but also have significantly lower faults per migration, approximately 0.05, or 20 migrations vs a single fault due to severe thrashes. This is explained by scarce and distributed successive data requests in GESUMMV. 

\subsection{Thrashing and Escalated Costs} \label{sec:thrashingcost}
Applications in Category III encounter thrashes, and their frequency increase with the degree of oversubscription. Here we study impacts of thrashing and associated escalated costs. 

\begin{figure}
    \centering
    \subfloat[][Eviction-to-Migration Ratio]{\includegraphics[width=0.80\linewidth]{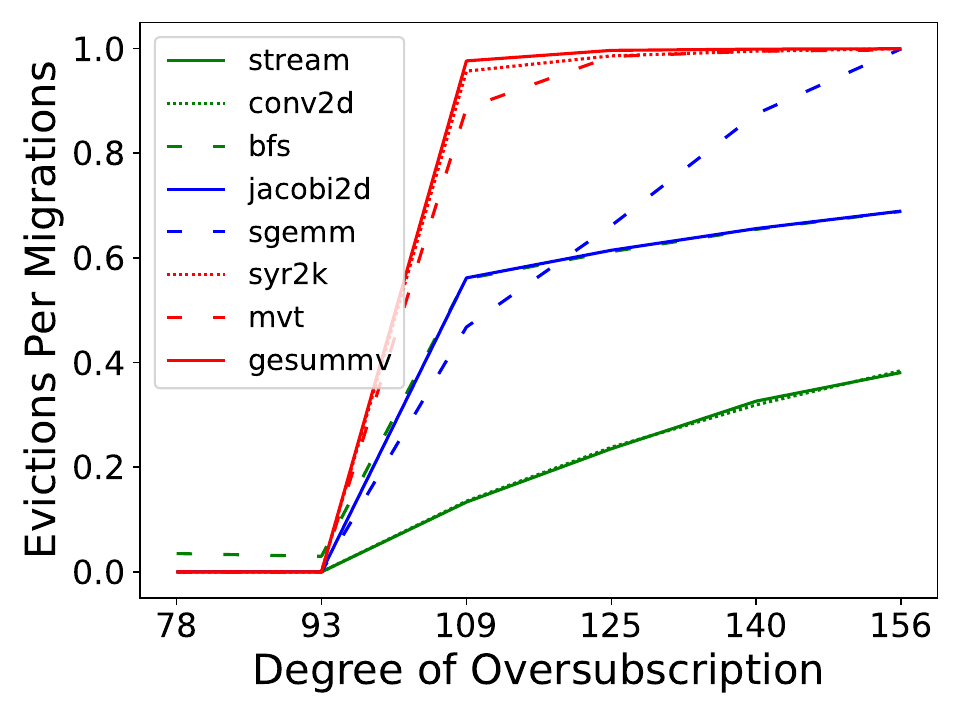}}\\
    \subfloat[][Escalated Migrations]{\includegraphics[width=0.80\linewidth]{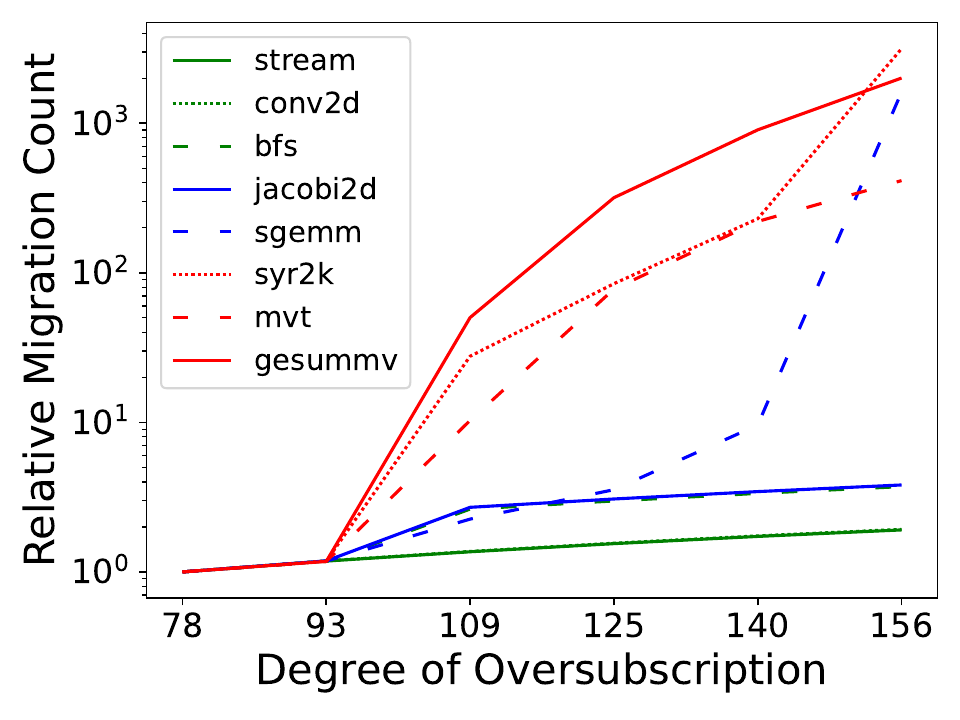}}
    \vspace{-5pt}
    \caption{Performance impacts of thrashing. The number of migrations are normalized to those at $DOS=78$}
    \vspace{-10pt}
    \label{fig:mig-growth}
    \Description{Growth in number of evictions (a) and relative amount of migrations (b) as all applications are oversubscribed.}
\end{figure}

Figure~\ref{fig:mig-growth} shows the increases of the eviction-to-migration ratio and migration counts with DOS. The eviction-to-migration ratio is 0 at $DOS < 100$ across the applications except BFS which algorithmically transfers data from the device to the host. The ratio quickly increases to 1 for applications in Category III, but only gradually increases for other applications, especially for application in Category I. A higher ratio corresponds to a higher cost per migration, and a ratio of 1 indicates each migration involves an eviction and thus doubled cost per migration.

The most severe performance degradation results from the order of magnitude increase in the migration counts. As DOS increases, the counts for applications in Category III  increase drastically by an order of magnitude or more, and increase exponentially for SGEMM and SYR2K once DOS reaches 140. In contrast, the counts for STREAM and Conv2d only increase linearly, doubling when DOS doubles. The count for Jacobi2d initially exhibits a jump and then proceeds to increase linearly. 
\section{Considerations and Discussions}
\label{sec:implication}

We discuss how application performance can be improved with SVM-aware algorithms and potential changes in SVM design. 

\subsection{SVM-Aware Algorithm Design}\label{sec:opt-alg}

Understanding the SVM design empowers us to identify the performance bottlenecks in existing algorithms and redesign them to optimize their interactions with SVM. In this section, we use SGEMM and Jacobi2d as case studies to demonstrate SVM-aware algorithm design. These case studies do not require changes to the current SVM design and parameters.

\begin{algorithm}
    \caption{Original Jacobi2d GPU Implementation}
    \label{alg:j2d_code1}
    \begin{algorithmic}
        \STATE GPU\_Kernel1:\
        \begin{ALC@g}
        \STATE $B[i,j] \gets 0.2 \times (A[i,j] + A[i-1,j] + A[i+1,j] + A[i,j-1] + A[i,j+1])$ 
        \end{ALC@g}

        \STATE GPU\_Kernel2:
        \begin{ALC@g}
        \STATE $A[i,j] \gets 0.2 \times (B[i,j] + B[i-1,j] + B[i+1,j] + B[i,j-1] + B[i,j+1])$ 
        \end{ALC@g}
    \end{algorithmic}
\end{algorithm}
\begin{algorithm}
    \caption{SVM-Aware Jacobi2d GPU Implementation}
    \label{alg:j2d_code2}
    \begin{algorithmic}
        \STATE GPU\_Kernel1:
        \begin{ALC@g}
        \STATE $B[i,j] \gets 0.2 \times (A[i,j] + A[i-1,j] + A[i+1,j] + A[i,j-1] + A[i,j+1])$ 
        \end{ALC@g}

        \STATE GPU\_Kernel2:
        \begin{ALC@g}
        \STATE $A[N-i,M-j] \gets 0.2 \times (B[N-i,M-j] + B[N-i-1,M-j] + B[N-i+1,j] + B[N-i,M-j-1] + B[N-i,M-j+1])$ 
        \end{ALC@g}
    \end{algorithmic}
\end{algorithm}

Jacobi2d's performance suffers in oversubscribed problem sizes due to an excessive amount of thrashing that is more than necessary. Jacobi2d iterates over two consecutive GPU kernels involving two matrices, each performing a partial convolution over one matrix to update the other, shown in Algorithm~\ref{alg:j2d_code1}. Both kernels traverse matrix data elements in the same manner, i.e., from the first to the last row and from left to right within each row. Under oversubscription, the first kernel evicts the first rows needed at the beginning of the second kernel execution, while the second kernel then progressively evicts the later rows needed soon by itself. Consequently, each range undergoes premature eviction  and thrashing. 

\begin{figure}
    \centering
    \includegraphics[width=0.80\linewidth]{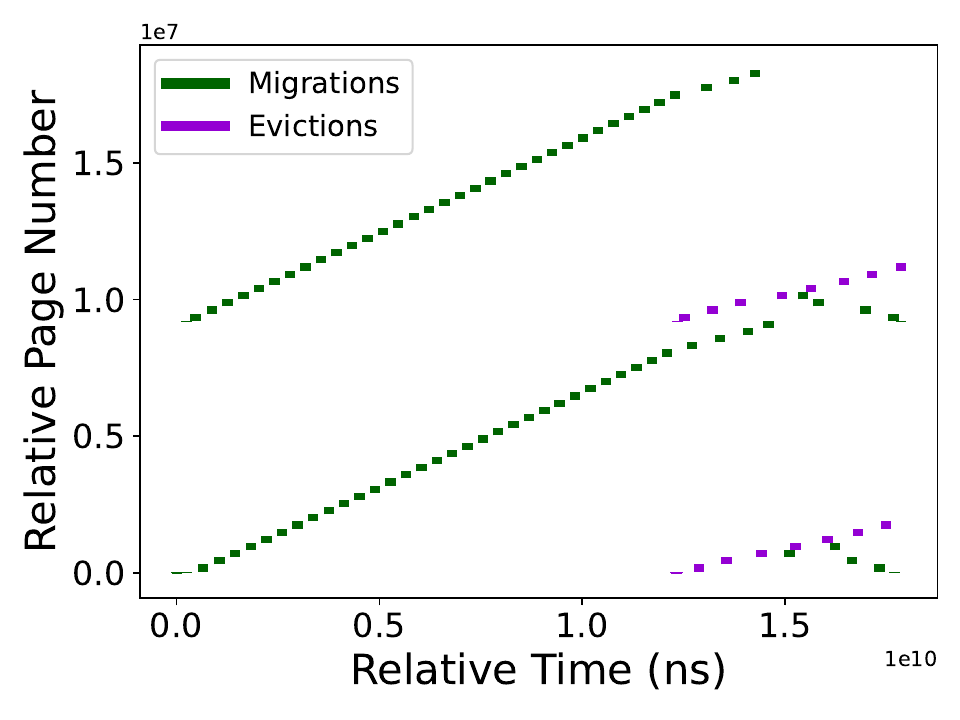}
    \vspace{-7pt}
    \caption{Timeline of migrations and evictions of the SVM-aware Jacobi2d implementation for small oversubscription. Compared to Figure \ref{subfig:j2d-ap}, thrashing is reduced significantly.}
    \vspace{-12pt}
    \label{fig:j2dopt-ap}
    \Description{Comparative timeline of migrations and evictions for SVM-Aware Jacobi2d application.}
\end{figure}

In the SVM-aware algorithm, we adjust the traversal order of the second kernel, i.e., from the last to the first row, and from right to left within each row, as in Algorithm~\ref{alg:j2d_code2}. Such an adjustment allows the next kernel to fully reuse data residing on the GPU memory in the same iteration  and across the consecutive iterations. Figure \ref{fig:j2dopt-ap} shows the migration and eviction timeline for the SVM-aware Jacobi2d implementation. Compared to the timeline of the original implementation, shown in Figure \ref{subfig:j2d-ap}, we see significantly less evictions needed for the same computation. 

SGEMM's performance drops to 0\% of relative performance. While the source code is unavailable to the public, its migration and eviction profiles in Figure~\ref{subfig:sgemm-ap} indicate that SGEMM first simultaneously migrates the entire allocation of each factor matrix, and then computes the product matrix row by row. As the number of computed rows of the product matrix is large enough to fill the GPU memory, new rows to be computed cause the eviction of factor matrix elements currently needed in computation. At this point, computation halts to re-migrate the newly evicted factor elements, which causes the eviction of the remaining factor elements. This chain of thrashing over factor matrix elements continues until the computed product rows become the least recently faulted and are evicted to make space for the new product rows. We speculate that SGEMM computes the total sum for a single product element at once by using an entire row of one factor to multiple an entire column of the other. 

\begin{figure}
    \centering
    \subfloat[][SGEMM]{\label{subfig:sgemm-ap-100}\includegraphics[width=0.9\linewidth]{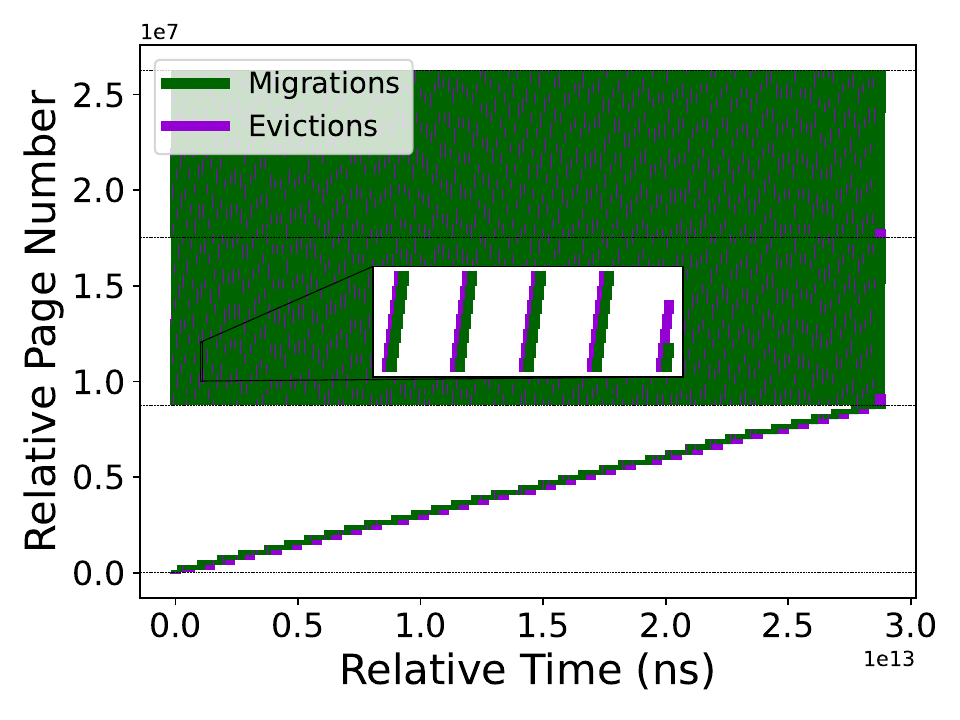}}
    \\
    \vspace{-12pt}
    \subfloat[][SGEMM-svm-aware]{\label{subfig:sgemm-aware-ap-100}\includegraphics[width=0.9\linewidth]{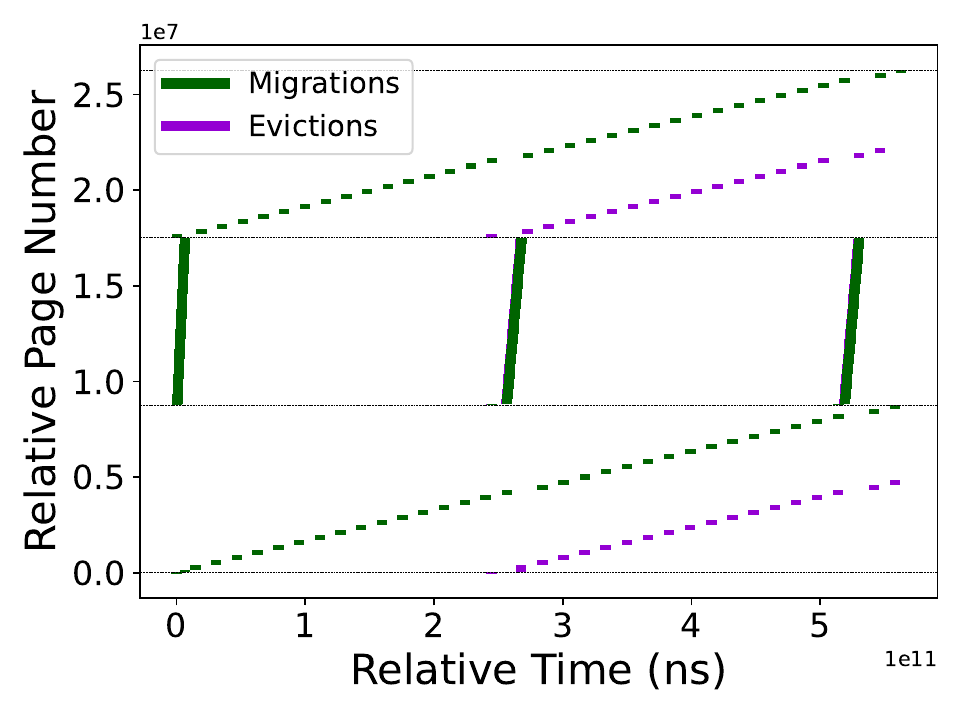}}
    \vspace{-3pt}
    \caption{Migration and eviction profiles of SGEMM and SGEMM-svm-aware at $DOS=156$.}
    \label{fig:sgemm-ap-100}
    \vspace{-10pt}
    \Description{Comparative migration and eviction timelines for both SGEMM and the SVM-Aware SGEMM at max problem sizes.}
\end{figure}

SGEMM is not scalable to support large problem sizes. For large problem sizes that cannot simultaneously fit both factor matrices in the GPU memory, the ranges of both factors are in a constant state of thrashing, depicted in Figure \ref{subfig:sgemm-ap-100}.

We design a naive but SVM-aware GPU SGEMM implementation \textit{SGEMM-svm-aware}, solely for the purpose of demonstrating the benefits of SVM-awareness instead of ultimate performance optimization. SGEMM-svm-aware migrates the entire column factor matrix to the GPU, and assigns a GPU thread block with a factor sub-matrix and the corresponding product sub-matrix to compute partial sums of the product elements. The total sums are carried out across the thread blocks. Computation only needs a chuck of rows for the row factor and the product at a time, and progresses over the rows. SGEMM-svm-aware significantly reduces the amount of thrashing, as shown in Figure \ref{subfig:sgemm-aware-ap-100}. Specifically, only one factor matrix (middle allocation) experiences thrashing twice, and the others encounter permanent evictions only.
 
\begin{figure}[t]
    \centering
    \includegraphics[width=0.80\linewidth]{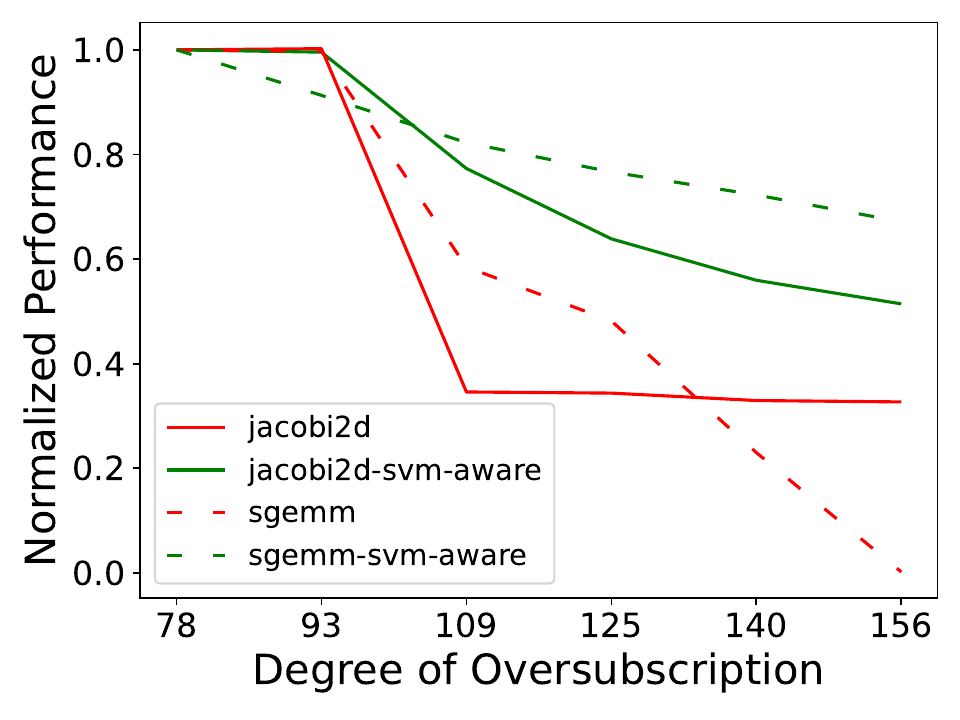}
    \vspace{-7pt}
    \caption{Comparison between the performance degradation of the original and SVM-aware Jacobi2d implementations.}
    \label{fig:aware-norm-comp}
    \vspace{-12pt}
    \Description{Comparative performance trends between original and improved SVM-Aware applications.}
\end{figure}

Figure \ref{fig:aware-norm-comp} presents the overall performance improvement using the SVM-aware algorithms. These algorithms show advantages over the counterparts in two aspects: preventing sudden drops in performance and elevating the lower performance limits. The SVM-aware Jacobi2d improves the performance at $DOS=109$ by more than 2X and improves the lower limit by 1.5X. SGEMM-svm-aware achieves a performance of 0.75 at $DOS=156$, in comparison to near zero with the counterpart, resulting in a speedup by several orders of magnitude. We recognize this algorithm only scales to $DOS \approx 300$, and a different algorithm is needed if DOS grows past.   

\subsection{Driver Design Considerations}

We focus on eviction-related implementation and design as eviction causes the most severe performance degradation. We first discuss implementations without changing the current design of policies and then discuss design alternatives. 

\textbf{Parallel Implementation.} As shown in Figure~\ref{fig:migrate_timeline}, eviction is a synchronous event in the SVM driver, blocking  migration of data needed in computation. Upon a migration request, SVM performs one eviction and then attempts the allocation again. In case of success, migration follows. Otherwise, SVM continues to perform another eviction and re-attempt until there is sufficient space.  A migration involving one eviction results in doubled cost. A migration involving a series of evictions and allocation attempts  not only increases the cost by multi-fold, but is error-prone. We observe in our experimental data that some allocations could take up to 15 seconds (a timeout in SVM for eviction; all these timeouts are removed in the figures). 

Parallel implementations explore overlapping eviction and migration using multithreading, i.e., one thread per eviction/migration. Parallel implementations are feasible, as the driver has the knowledge of ranges to be evicted and migrated. Such parallelization, though involving locks, is expected to improve performance, especially for migrations involving multiple evictions. The current\textit{} SVM already includes mechanisms for quick rollbacks in case of allocation failures. It can be adapted to support multithreading. 

\textbf{Eviction Policy. } The Least Recently Faulted (LRF) policy evicts the range that has been migrated to the GPU earliest, ignorant of whether the range has been actively used or will be needed in the future while residing on the GPU. Consequently, it tends to evict the most intensely reused ``hot'' data from the GPU for SGEMM and alike, drastically degrading their performances.

A commonly used policy is the Least Recently Used (LRU). However, LRU requires the driver to timestamp page access on the GPU, which is too costly. Instead, we can explore simplified versions such as the Clock algorithm~\cite{jiang2005}, which  groups the ranges into two types: hot and cold.  To avoid the prohibitive communication overhead between the device and the driver, the device could keep a copy of the range metadata and make the eviction decision. It would be trivial for the device to communicate the decision to the driver using existing communications. 

\textbf{Granularity.} The large range sizes exacerbate  eviction and the cost. A single data access missed on the GPU triggers the migration of the entire range. \textit{Such behavior is the extreme case of aggressive prefetching}, which rapidly fills the GPU memory and leads to oversubscription and frequent thrashing. 

Reducing the migration granularity or enabling adaptive ``prefetching'' would significantly benefit applications in Category III. Adjusting the range size would benefit applications with sparse or non-linear accesses and improve application SVM and alike.  Instead of immediately migrating an entire range after one data request, the driver could migrate the range only after an access count has been reached or a percentage of data has been requested, similar to access counter based and density-based prefetching in UVM~\cite{he2008}. 

\textbf{Zero-Copy instead of Demand Paging. } The SVM driver supports both demand paging and zero-copy for accessing unified memory. With zero-copy, the device accesses data residing on the host at the cache line granularity. Each zero-copy incurs a large latency across the interconnect in comparison to local memory accesses. However, zero-copy is expected to benefit applications that experience severe thrashing under demand paging~\cite{emogi}. For such applications, the driver can allocate a portion of data in GPU memory for optimal utilization and allocate the remaining data on the host, accessing them via zero-copy.
\section{Related Work}
\label{sec:related}
Prior works studying the performance of unified memory systems can be looked at in two categories: (1) application-level analysis that focuses on the performance of select GPU applications utilizing different memory configurations and (2) system-level analysis that focuses on breaking down the components and features of unified memory implementations then directly profiling their cost. Most prior work exists in  UVM with some work beginning to break into various HMM implementations, including  SVM. 

\textbf{Application-Level Analysis} works focus on the performance of GPU applications in unified memory models. These works span comparing performance between unified memory and programmer-managed GPU memory, analyzing the impact of various hardware on unified memory, identifying key application features impacting unified memory, and attempting to mitigate the performance loss of unified memory. Several works exist examining the performance differences between non-unified and unified memory across CUDA applications with combinations of prefetching \& oversubscription \cite{Knap2019-fl,landaverdeHPEC,nadalTPDS,Yu2019b}. Chien et al. further examine the impact of memory hints and advice the programmer can provide in CUDA on application performance \cite{chienMCHPC}. Xu et al. introduce a framework that relates application features with UVM hints, removing the need for programmers to decide which hints to use~\cite{xuUVMhints}. Gayatri et al. study the impact of Address Translation Services (ATS) for Power systems on unified memory~\cite{gayatriPMBS}. All works discussed so far are in NVIDIA's ecosystem. To our knowledge,  little work has examined unified memory in AMD's ecosystem except~\cite{jin-ipdpsw22}, which evaluates the performance of AMD GPU applications across user-managed memory, zero-copy remote access, and non-xnack managed memory.

\textbf{System-Level Analysis} works are the closest in relation to our work. Kim et al. identify the fault batching behavior of UVM, provide initial results on the relationship between batch size and batch cost, and propose a thread oversubscription technique for the GPU to mitigate large amounts of small batches \cite{kim-asplos20}. Allen and Ge explore the driver-level impact of prefetching in UVM, provide initial insight on distinct components of UVM batches, and highlight how application access patterns influence GPU performance \cite{allen-ipdps21}. Allen and Ge further explore the components associated with servicing on-demand faults in UVM and how batch features influence the cost of various components \cite{allen-sc21}. Our work explores the behavioral features of AMD's unified memory, identifies the unique interplay between vendor UM and HMM, and analyzes the cost distribution in servicing on-demand page faults in an HMM-based unified memory.

\textbf{Optimization} works implement novel changes to unified memory drivers in order to lessen their performance cost. Chang et al. present an adaptive page migration scheme that enhances NVIDIA's UVM driver to dynamically migrate pages for irregular applications~\cite{chang21}. Ganguly et al. combine hardware prefetching and a novel pre-eviction policy to improve performance of oversubscribed UVM applications~\cite{ganguly19}. Li et al. propose a framework with proactive eviction, memory-aware throttling, and capacity compression to improve the performance of GPU oversubscription~\cite{li19}. Such work is largely driven by application-level insights, while ours delves into the driver-level design and can be used in combination to optimize performance further.
\section{Conclusion and Future Work}
\label{sec:conclusion}
In this work, we have examined the SVM design and its interface with the Linux kernel's HMM, and studied the performance impacts on a diverse set of GPU workloads. We find that the SVM design, especially the SVM ranges, are different from UVM. This design results in significant performance degradation for certain workloads under oversubscription. We show this performance degradation can be mitigated in part with  SVM aware algorithms. We further discuss SVM design changes that would reduce the performance issues.

There is ample work to be done for SVM and in the larger field of unified memory. We have intentionally focused on  demand paging from host to device,  dismissing the directions from device to device and from device to host. The experiments presented throughout this paper solely use memory allocated through \texttt{hipMallocManaged()}. SVM/HMM allows other memory allocation such as \texttt{malloc()} and pinning on the host. Some combinations of various allocations may have observable performance effects.

Extending the work to different systems is another possibility. AMD's next generation of HPC processors, MI300~\cite{mi300}, are in the form of Accelerated Processing Units (APU) with tightly coupled CPU cores and GCDs in a shared processing unit. All CPU cores and GCDs share the same physical memory. The implications this architecture will have on SVM are unknown. Further analyzing HMM and its interface supporting various GPU devices is another avenue. At the start of this work, SVM was the only robust unified memory system utilizing Linux's HMM component. UVM has since been extended to support HMM~\cite{nvidia-hmm}. 

Lastly, this work serves as a first step in analyzing and improving the current SVM implementation. The SVM-aware algorithm designs can serve as a base for other applications to improve performance under SVM. The discussed driver-level changes can be attempted and tested on actual systems. 

\begin{acks}
This research is partially supported by U.S. National Science Foundation under Grants CCF-1942182. This work was produced under the auspices of the U.S. Department of Energy by Lawrence Livermore National Laboratory under Contract DE-AC52-07NA27344.
\end{acks}

\bibliographystyle{ACM-Reference-Format}
\bibliography{paper}


\begin{thebibliography}{46}


\ifx \showCODEN    \undefined \def \showCODEN     #1{\unskip}     \fi
\ifx \showDOI      \undefined \def \showDOI       #1{#1}\fi
\ifx \showISBNx    \undefined \def \showISBNx     #1{\unskip}     \fi
\ifx \showISBNxiii \undefined \def \showISBNxiii  #1{\unskip}     \fi
\ifx \showISSN     \undefined \def \showISSN      #1{\unskip}     \fi
\ifx \showLCCN     \undefined \def \showLCCN      #1{\unskip}     \fi
\ifx \shownote     \undefined \def \shownote      #1{#1}          \fi
\ifx \showarticletitle \undefined \def \showarticletitle #1{#1}   \fi
\ifx \showURL      \undefined \def \showURL       {\relax}        \fi
\providecommand\bibfield[2]{#2}
\providecommand\bibinfo[2]{#2}
\providecommand\natexlab[1]{#1}
\providecommand\showeprint[2][]{arXiv:#2}

\bibitem[Allen and Ge(2021a)]%
        {allen-ipdps21}
\bibfield{author}{\bibinfo{person}{Tyler Allen} {and} \bibinfo{person}{Rong Ge}.} \bibinfo{year}{2021}\natexlab{a}.
\newblock \showarticletitle{Demystifying {GPU UVM} Cost with Deep Runtime and Workload Analysis}. In \bibinfo{booktitle}{\emph{2021 IEEE International Parallel and Distributed Processing Symposium (IPDPS)}}. \bibinfo{pages}{141--150}.
\newblock


\bibitem[Allen and Ge(2021b)]%
        {allen-sc21}
\bibfield{author}{\bibinfo{person}{Tyler Allen} {and} \bibinfo{person}{Rong Ge}.} \bibinfo{year}{2021}\natexlab{b}.
\newblock \showarticletitle{In-Depth Analyses of Unified Virtual Memory System for {GPU} Accelerated Computing}. In \bibinfo{booktitle}{\emph{Proceedings of the International Conference for High Performance Computing, Networking, Storage and Analysis}} (St. Louis, Missouri) \emph{(\bibinfo{series}{SC '21})}. \bibinfo{publisher}{Association for Computing Machinery}, \bibinfo{address}{New York, NY, USA}, Article \bibinfo{articleno}{64}, \bibinfo{numpages}{15}~pages.
\newblock
\showISBNx{9781450384421}


\bibitem[{AMD}(2023)]%
        {rocm}
\bibfield{author}{\bibinfo{person}{{AMD}}.} \bibinfo{year}{2023}\natexlab{}.
\newblock \bibinfo{title}{{AMD} {ROC}m™ documentation}.
\newblock
\newblock
\urldef\tempurl%
\url{https://rocm.docs.amd.com/en/latest/}
\showURL{%
\tempurl}


\bibitem[AMD(2023)]%
        {rocblas}
\bibfield{author}{\bibinfo{person}{AMD}.} \bibinfo{year}{2023}\natexlab{}.
\newblock \bibinfo{title}{roc{BLAS} Documentation}.
\newblock
\newblock
\urldef\tempurl%
\url{https://rocblas.readthedocs.io/en/master/index.html}
\showURL{%
\tempurl}


\bibitem[AMD(2024)]%
        {mi300}
\bibfield{author}{\bibinfo{person}{AMD}.} \bibinfo{year}{2024}\natexlab{}.
\newblock \bibinfo{title}{{AMD} Instinct™ {MI300} Series Accelerators}.
\newblock
\newblock
\urldef\tempurl%
\url{https://www.amd.com/en/products/accelerators/instinct/mi300.html}
\showURL{%
\tempurl}


\bibitem[Beckingsale et~al\mbox{.}(2019)]%
        {raja}
\bibfield{author}{\bibinfo{person}{David~A. Beckingsale}, \bibinfo{person}{Jason Burmark}, \bibinfo{person}{Rich Hornung}, {et~al\mbox{.}}} \bibinfo{year}{2019}\natexlab{}.
\newblock \showarticletitle{{RAJA}: Portable Performance for Large-Scale Scientific Applications}. In \bibinfo{booktitle}{\emph{2019 IEEE/ACM International Workshop on Performance, Portability and Productivity in HPC (P3HPC)}}. \bibinfo{pages}{71--81}.
\newblock


\bibitem[Ben-Nun et~al\mbox{.}(2022)]%
        {tal2022}
\bibfield{author}{\bibinfo{person}{Tal Ben-Nun}, \bibinfo{person}{Linus Groner}, \bibinfo{person}{Florian Deconinck}, {et~al\mbox{.}}} \bibinfo{year}{2022}\natexlab{}.
\newblock \showarticletitle{Productive performance engineering for weather and climate modeling with Python}. In \bibinfo{booktitle}{\emph{Proceedings of the International Conference on High Performance Computing, Networking, Storage and Analysis}} (Dallas, Texas) \emph{(\bibinfo{series}{SC '22})}. \bibinfo{publisher}{IEEE Press}, Article \bibinfo{articleno}{73}, \bibinfo{numpages}{14}~pages.
\newblock
\showISBNx{9784665454445}


\bibitem[Brown et~al\mbox{.}(2020)]%
        {brown2020}
\bibfield{author}{\bibinfo{person}{Tom Brown}, \bibinfo{person}{Benjamin Mann}, \bibinfo{person}{Nick Ryder}, {et~al\mbox{.}}} \bibinfo{year}{2020}\natexlab{}.
\newblock \showarticletitle{Language models are few-shot learners}.
\newblock \bibinfo{journal}{\emph{Advances in neural information processing systems}}  \bibinfo{volume}{33} (\bibinfo{year}{2020}), \bibinfo{pages}{1877--1901}.
\newblock


\bibitem[Carter~Edwards et~al\mbox{.}(2014)]%
        {kokkos}
\bibfield{author}{\bibinfo{person}{H. Carter~Edwards}, \bibinfo{person}{Christian~R. Trott}, {and} \bibinfo{person}{Daniel Sunderland}.} \bibinfo{year}{2014}\natexlab{}.
\newblock \showarticletitle{Kokkos}.
\newblock \bibinfo{journal}{\emph{J. Parallel Distrib. Comput.}} \bibinfo{volume}{74}, \bibinfo{number}{12} (\bibinfo{date}{Dec.} \bibinfo{year}{2014}), \bibinfo{pages}{3202–3216}.
\newblock
\showISSN{0743-7315}


\bibitem[Chang et~al\mbox{.}(2021)]%
        {chang21}
\bibfield{author}{\bibinfo{person}{Chia-Hao Chang}, \bibinfo{person}{Adithya Kumar}, {and} \bibinfo{person}{Anand Sivasubramaniam}.} \bibinfo{year}{2021}\natexlab{}.
\newblock \showarticletitle{To move or not to move? page migration for irregular applications in over-subscribed GPU memory systems with DynaMap}. In \bibinfo{booktitle}{\emph{Proceedings of the 14th ACM International Conference on Systems and Storage}} (Haifa, Israel) \emph{(\bibinfo{series}{SYSTOR '21})}. \bibinfo{publisher}{Association for Computing Machinery}, \bibinfo{address}{New York, NY, USA}, Article \bibinfo{articleno}{1}, \bibinfo{numpages}{12}~pages.
\newblock
\showISBNx{9781450383981}


\bibitem[Chien et~al\mbox{.}(2019)]%
        {chienMCHPC}
\bibfield{author}{\bibinfo{person}{Steven Chien}, \bibinfo{person}{Ivy Peng}, {and} \bibinfo{person}{Stefano Markidis}.} \bibinfo{year}{2019}\natexlab{}.
\newblock \showarticletitle{Performance Evaluation of Advanced Features in {CUDA} Unified Memory}. In \bibinfo{booktitle}{\emph{2019 IEEE/ACM Workshop on Memory Centric High Performance Computing (MCHPC)}}. \bibinfo{pages}{50--57}.
\newblock


\bibitem[Choi et~al\mbox{.}(2022)]%
        {Choi2022ImprovingOG}
\bibfield{author}{\bibinfo{person}{Jake Choi}, \bibinfo{person}{Heon~Young Yeom}, {and} \bibinfo{person}{Yoonhee Kim}.} \bibinfo{year}{2022}\natexlab{}.
\newblock \showarticletitle{Improving Oversubscribed GPU Memory Performance in the PyTorch Framework}.
\newblock \bibinfo{journal}{\emph{Cluster Computing}}  \bibinfo{volume}{26} (\bibinfo{year}{2022}), \bibinfo{pages}{2835 -- 2850}.
\newblock
\urldef\tempurl%
\url{https://api.semanticscholar.org/CorpusID:253492267}
\showURL{%
\tempurl}


\bibitem[Ganguly et~al\mbox{.}(2019)]%
        {ganguly19}
\bibfield{author}{\bibinfo{person}{Debashis Ganguly}, \bibinfo{person}{Ziyu Zhang}, \bibinfo{person}{Jun Yang}, {and} \bibinfo{person}{Rami Melhem}.} \bibinfo{year}{2019}\natexlab{}.
\newblock \showarticletitle{Interplay between hardware prefetcher and page eviction policy in CPU-GPU unified virtual memory}. In \bibinfo{booktitle}{\emph{Proceedings of the 46th International Symposium on Computer Architecture}} (Phoenix, Arizona) \emph{(\bibinfo{series}{ISCA '19})}. \bibinfo{publisher}{Association for Computing Machinery}, \bibinfo{address}{New York, NY, USA}, \bibinfo{pages}{224–235}.
\newblock
\showISBNx{9781450366694}


\bibitem[Gayatri et~al\mbox{.}(2019)]%
        {gayatriPMBS}
\bibfield{author}{\bibinfo{person}{Rahulkumar Gayatri}, \bibinfo{person}{Kevin Gott}, {and} \bibinfo{person}{Jack Deslippe}.} \bibinfo{year}{2019}\natexlab{}.
\newblock \showarticletitle{Comparing Managed Memory and {ATS} with and without Prefetching on {NVIDIA} Volta GPUs}. In \bibinfo{booktitle}{\emph{2019 IEEE/ACM Performance Modeling, Benchmarking and Simulation of High Performance Computer Systems (PMBS)}}. \bibinfo{pages}{41--46}.
\newblock


\bibitem[He et~al\mbox{.}(2008)]%
        {he2008}
\bibfield{author}{\bibinfo{person}{Yanxiang He}, \bibinfo{person}{Shaohua Wan}, \bibinfo{person}{Naixue Xiong}, {and} \bibinfo{person}{Jong~Hyuk Park}.} \bibinfo{year}{2008}\natexlab{}.
\newblock \showarticletitle{A New Prefetching Strategy Based on Access Density in Linux}. In \bibinfo{booktitle}{\emph{International Symposium on Computer Science and its Applications}}. \bibinfo{pages}{22--27}.
\newblock


\bibitem[Heroux et~al\mbox{.}(2005)]%
        {trilinos}
\bibfield{author}{\bibinfo{person}{Michael~A. Heroux}, \bibinfo{person}{Roscoe~A. Bartlett}, \bibinfo{person}{Vicki~E. Howle}, {et~al\mbox{.}}} \bibinfo{year}{2005}\natexlab{}.
\newblock \showarticletitle{An Overview of the Trilinos Project}.
\newblock \bibinfo{journal}{\emph{ACM Trans. Math. Softw.}} \bibinfo{volume}{31}, \bibinfo{number}{3} (\bibinfo{date}{Sept.} \bibinfo{year}{2005}), \bibinfo{pages}{397–423}.
\newblock
\showISSN{0098-3500}


\bibitem[Hornung and Hones(2017)]%
        {rajaPerf}
\bibfield{author}{\bibinfo{person}{Richard~D. Hornung} {and} \bibinfo{person}{Holger~E. Hones}.} \bibinfo{year}{2017}\natexlab{}.
\newblock \bibinfo{title}{{RAJA} Performance Suite}.
\newblock \bibinfo{howpublished}{[Computer Software] \url{https://doi.org/10.11578/dc.20201001.36}}.
\newblock
\urldef\tempurl%
\url{https://doi.org/10.11578/dc.20201001.36}
\showDOI{\tempurl}


\bibitem[Hubbard et~al\mbox{.}(2023)]%
        {nvidia-hmm}
\bibfield{author}{\bibinfo{person}{John Hubbard}, \bibinfo{person}{Gonzalo Brito}, \bibinfo{person}{Chirayu Garg}, {et~al\mbox{.}}} \bibinfo{year}{2023}\natexlab{}.
\newblock \bibinfo{title}{Simplifying {GPU} application development with heterogeneous memory management}.
\newblock
\newblock
\urldef\tempurl%
\url{https://developer.nvidia.com/blog/simplifying-gpu-application-development-with-heterogeneous-memory-management/}
\showURL{%
\tempurl}


\bibitem[Hubbard and Glisee(2017)]%
        {hmm}
\bibfield{author}{\bibinfo{person}{John Hubbard} {and} \bibinfo{person}{Jerome Glisee}.} \bibinfo{year}{2017}\natexlab{}.
\newblock \bibinfo{title}{{GPUs}: {HMM}: Heterogeneous Memory Management}.
\newblock
\newblock
\urldef\tempurl%
\url{https://www.redhat.com/files/summit/session-assets/2017/S104078-hubbard.pdf}
\showURL{%
\tempurl}


\bibitem[Jiang et~al\mbox{.}(2005)]%
        {jiang2005}
\bibfield{author}{\bibinfo{person}{Song Jiang}, \bibinfo{person}{Feng Chen}, {and} \bibinfo{person}{Xiaodong Zhang}.} \bibinfo{year}{2005}\natexlab{}.
\newblock \showarticletitle{{CLOCK-Pro}: An Effective Improvement of the {CLOCK} Replacement}. In \bibinfo{booktitle}{\emph{2005 USENIX Annual Technical Conference (USENIX ATC 05)}}. \bibinfo{publisher}{USENIX Association}, \bibinfo{address}{Anaheim, CA}.
\newblock


\bibitem[Jin and Vetter(2022)]%
        {jin-ipdpsw22}
\bibfield{author}{\bibinfo{person}{Zheming Jin} {and} \bibinfo{person}{Jeffrey~S. Vetter}.} \bibinfo{year}{2022}\natexlab{}.
\newblock \showarticletitle{Evaluating Unified Memory Performance in {HIP}}. In \bibinfo{booktitle}{\emph{2022 IEEE International Parallel and Distributed Processing Symposium Workshops (IPDPSW)}}. \bibinfo{pages}{562--568}.
\newblock


\bibitem[Jung et~al\mbox{.}(2023)]%
        {Jung2023DeepUM}
\bibfield{author}{\bibinfo{person}{Jaehoon Jung}, \bibinfo{person}{Jinpyo Kim}, {and} \bibinfo{person}{Jaejin Lee}.} \bibinfo{year}{2023}\natexlab{}.
\newblock \showarticletitle{DeepUM: Tensor Migration and Prefetching in Unified Memory}. In \bibinfo{booktitle}{\emph{Proceedings of the 28th ACM International Conference on Architectural Support for Programming Languages and Operating Systems, Volume 2}} (Vancouver, BC, Canada) \emph{(\bibinfo{series}{ASPLOS 2023})}. \bibinfo{publisher}{Association for Computing Machinery}, \bibinfo{address}{New York, NY, USA}, \bibinfo{pages}{207–221}.
\newblock
\showISBNx{9781450399166}


\bibitem[{Khronos Group}(2023)]%
        {opencl}
\bibfield{author}{\bibinfo{person}{{Khronos Group}}.} \bibinfo{year}{2023}\natexlab{}.
\newblock \bibinfo{title}{Open Standard for Parallel Programming of Heterogeneous Systems}.
\newblock
\newblock
\urldef\tempurl%
\url{https://www.khronos.org/api/opencl}
\showURL{%
\tempurl}


\bibitem[Kim et~al\mbox{.}(2020)]%
        {kim-asplos20}
\bibfield{author}{\bibinfo{person}{Hyojong Kim}, \bibinfo{person}{Jaewoong Sim}, \bibinfo{person}{Prasun Gera}, {et~al\mbox{.}}} \bibinfo{year}{2020}\natexlab{}.
\newblock \showarticletitle{Batch-Aware Unified Memory Management in {GPUs} for Irregular Workloads}. In \bibinfo{booktitle}{\emph{Proceedings of the Twenty-Fifth International Conference on Architectural Support for Programming Languages and Operating Systems}} (Lausanne, Switzerland) \emph{(\bibinfo{series}{ASPLOS '20})}. \bibinfo{publisher}{Association for Computing Machinery}, \bibinfo{address}{New York, NY, USA}, \bibinfo{pages}{1357–1370}.
\newblock
\showISBNx{9781450371025}


\bibitem[Knap and Czarnul(2019)]%
        {Knap2019-fl}
\bibfield{author}{\bibinfo{person}{Marcin Knap} {and} \bibinfo{person}{Pawe{\l} Czarnul}.} \bibinfo{year}{2019}\natexlab{}.
\newblock \showarticletitle{Performance evaluation of Unified Memory with prefetching and oversubscription for selected parallel {CUDA} applications on {NVIDIA} Pascal and Volta {GPUs}}.
\newblock \bibinfo{journal}{\emph{The Journal of Supercomputing}} \bibinfo{volume}{75}, \bibinfo{number}{11} (\bibinfo{date}{Nov.} \bibinfo{year}{2019}), \bibinfo{pages}{7625--7645}.
\newblock


\bibitem[Landaverde et~al\mbox{.}(2014)]%
        {landaverdeHPEC}
\bibfield{author}{\bibinfo{person}{Raphael Landaverde}, \bibinfo{person}{Tiansheng Zhang}, \bibinfo{person}{Ayse~K. Coskun}, {and} \bibinfo{person}{Martin Herbordt}.} \bibinfo{year}{2014}\natexlab{}.
\newblock \showarticletitle{An investigation of Unified Memory Access performance in {CUDA}}. In \bibinfo{booktitle}{\emph{2014 IEEE High Performance Extreme Computing Conference (HPEC)}}. \bibinfo{pages}{1--6}.
\newblock


\bibitem[{Lawrence Livermore National Lab}(2022)]%
        {llnl-tioga}
\bibfield{author}{\bibinfo{person}{{Lawrence Livermore National Lab}}.} \bibinfo{year}{2022}\natexlab{}.
\newblock \bibinfo{title}{Tioga}.
\newblock
\newblock
\urldef\tempurl%
\url{https://hpc.llnl.gov/hardware/compute-platforms/tioga}
\showURL{%
\tempurl}


\bibitem[{Lawrence Livermore National Lab}(2023)]%
        {llnl-elcap}
\bibfield{author}{\bibinfo{person}{{Lawrence Livermore National Lab}}.} \bibinfo{year}{2023}\natexlab{}.
\newblock \bibinfo{title}{El Capitan: Preparing for {NNSA}’s first exascale machine}.
\newblock
\newblock
\urldef\tempurl%
\url{https://asc.llnl.gov/exascale/el-capitan}
\showURL{%
\tempurl}


\bibitem[Le~Scao et~al\mbox{.}(2023)]%
        {bloom}
\bibfield{author}{\bibinfo{person}{Teven Le~Scao}, \bibinfo{person}{Angela Fan}, \bibinfo{person}{Christopher Akiki}, {et~al\mbox{.}}} \bibinfo{year}{2023}\natexlab{}.
\newblock \bibinfo{title}{BLOOM: A 176B-Parameter Open-Access Multilingual Language Model}.
\newblock
\newblock
\showeprint[arxiv]{2211.05100}~[cs.CL]


\bibitem[Le\'{o}n et~al\mbox{.}(2020)]%
        {toss}
\bibfield{author}{\bibinfo{person}{Edgar~A. Le\'{o}n}, \bibinfo{person}{Trent D'Hooge}, \bibinfo{person}{Nathan Hanford}, {et~al\mbox{.}}} \bibinfo{year}{2020}\natexlab{}.
\newblock \showarticletitle{TOSS-2020: A Commodity Software Stack for HPC}. In \bibinfo{booktitle}{\emph{SC20: International Conference for High Performance Computing, Networking, Storage and Analysis}} (Atlanta, Georgia) \emph{(\bibinfo{series}{SC '20})}. \bibinfo{publisher}{IEEE Press}, Article \bibinfo{articleno}{40}, \bibinfo{numpages}{15}~pages.
\newblock
\showISBNx{9781728199986}


\bibitem[Li et~al\mbox{.}(2019)]%
        {li19}
\bibfield{author}{\bibinfo{person}{Chen Li}, \bibinfo{person}{Rachata Ausavarungnirun}, \bibinfo{person}{Christopher~J. Rossbach}, {et~al\mbox{.}}} \bibinfo{year}{2019}\natexlab{}.
\newblock \showarticletitle{A Framework for Memory Oversubscription Management in Graphics Processing Units}. In \bibinfo{booktitle}{\emph{Proceedings of the Twenty-Fourth International Conference on Architectural Support for Programming Languages and Operating Systems}} (Providence, RI, USA) \emph{(\bibinfo{series}{ASPLOS '19})}. \bibinfo{publisher}{Association for Computing Machinery}, \bibinfo{address}{New York, NY, USA}, \bibinfo{pages}{49–63}.
\newblock
\showISBNx{9781450362405}


\bibitem[{Linux Kernel Development Community}(2023)]%
        {linux-hmm}
\bibfield{author}{\bibinfo{person}{{Linux Kernel Development Community}}.} \bibinfo{year}{2023}\natexlab{}.
\newblock \bibinfo{title}{Heterogeneous Memory Management ({HMM})}.
\newblock
\newblock
\urldef\tempurl%
\url{https://www.kernel.org/doc/html/latest/mm/hmm.html}
\showURL{%
\tempurl}


\bibitem[Loh et~al\mbox{.}(2023)]%
        {amdRetro}
\bibfield{author}{\bibinfo{person}{Gabriel~H. Loh}, \bibinfo{person}{Michael~J. Schulte}, \bibinfo{person}{Mike Ignatowski}, {et~al\mbox{.}}} \bibinfo{year}{2023}\natexlab{}.
\newblock \showarticletitle{A Research Retrospective on {AMD's} Exascale Computing Journey}. In \bibinfo{booktitle}{\emph{Proceedings of the 50th Annual International Symposium on Computer Architecture}} (Orlando, FL, USA) \emph{(\bibinfo{series}{ISCA '23})}. \bibinfo{publisher}{Association for Computing Machinery}, \bibinfo{address}{New York, NY, USA}, Article \bibinfo{articleno}{81}, \bibinfo{numpages}{14}~pages.
\newblock
\showISBNx{9798400700958}


\bibitem[Min et~al\mbox{.}(2021)]%
        {Min2021PyTorchDirectEG}
\bibfield{author}{\bibinfo{person}{Seungwon Min}, \bibinfo{person}{Kun Wu}, \bibinfo{person}{Sitao Huang}, {et~al\mbox{.}}} \bibinfo{year}{2021}\natexlab{}.
\newblock \showarticletitle{PyTorch-Direct: Enabling {GPU} Centric Data Access for Very Large Graph Neural Network Training with Irregular Accesses}.
\newblock \bibinfo{journal}{\emph{CoRR}}  \bibinfo{volume}{abs/2101.07956} (\bibinfo{year}{2021}).
\newblock
\showeprint[arXiv]{2101.07956}


\bibitem[Min et~al\mbox{.}(2020)]%
        {emogi}
\bibfield{author}{\bibinfo{person}{Seung~Won Min}, \bibinfo{person}{Vikram~Sharma Mailthody}, \bibinfo{person}{Zaid Qureshi}, {et~al\mbox{.}}} \bibinfo{year}{2020}\natexlab{}.
\newblock \showarticletitle{EMOGI: Efficient Memory-Access for out-of-Memory Graph-Traversal in GPUs}.
\newblock \bibinfo{journal}{\emph{Proc. VLDB Endow.}} \bibinfo{volume}{14}, \bibinfo{number}{2} (\bibinfo{date}{oct} \bibinfo{year}{2020}), \bibinfo{pages}{114–127}.
\newblock
\showISSN{2150-8097}


\bibitem[Nadal-Serrano and Lopez-Vallejo(2016)]%
        {nadalTPDS}
\bibfield{author}{\bibinfo{person}{Jose~M. Nadal-Serrano} {and} \bibinfo{person}{Marisa Lopez-Vallejo}.} \bibinfo{year}{2016}\natexlab{}.
\newblock \showarticletitle{A Performance Study of {CUDA UVM} versus Manual Optimizations in a Real-World Setup: Application to a Monte Carlo Wave-Particle Event-Based Interaction Model}.
\newblock \bibinfo{journal}{\emph{IEEE Transactions on Parallel and Distributed Systems}} \bibinfo{volume}{27}, \bibinfo{number}{6} (\bibinfo{year}{2016}), \bibinfo{pages}{1579--1588}.
\newblock


\bibitem[{Oak Ridge National Lab}(2022)]%
        {frontier}
\bibfield{author}{\bibinfo{person}{{Oak Ridge National Lab}}.} \bibinfo{year}{2022}\natexlab{}.
\newblock \bibinfo{title}{Frontier User Guide - {OLCF} User Documentation}.
\newblock
\newblock
\urldef\tempurl%
\url{https://docs.olcf.ornl.gov/systems/frontier\_user\_guide.html}
\showURL{%
\tempurl}


\bibitem[Pham et~al\mbox{.}(2023)]%
        {pham23}
\bibfield{author}{\bibinfo{person}{Minh Pham}, \bibinfo{person}{Yicheng Tu}, {and} \bibinfo{person}{Xiaoyi Lv}.} \bibinfo{year}{2023}\natexlab{}.
\newblock \showarticletitle{Accelerating BWA-MEM Read Mapping on GPUs}. In \bibinfo{booktitle}{\emph{Proceedings of the 37th International Conference on Supercomputing}} (Orlando, FL, USA) \emph{(\bibinfo{series}{ICS '23})}. \bibinfo{publisher}{Association for Computing Machinery}, \bibinfo{address}{New York, NY, USA}, \bibinfo{pages}{155–166}.
\newblock
\showISBNx{9798400700569}


\bibitem[Prasad et~al\mbox{.}(2005)]%
        {systemtap}
\bibfield{author}{\bibinfo{person}{Vara Prasad}, \bibinfo{person}{William Cohen}, \bibinfo{person}{FC Eigler}, {et~al\mbox{.}}} \bibinfo{year}{2005}\natexlab{}.
\newblock \showarticletitle{Locating system problems using dynamic instrumentation}. In \bibinfo{booktitle}{\emph{2005 Ottawa Linux Symposium}}. New York, NY: IEEE, \bibinfo{pages}{49--64}.
\newblock


\bibitem[Smith and James(2022)]%
        {mi200-arch}
\bibfield{author}{\bibinfo{person}{Alan Smith} {and} \bibinfo{person}{Norman James}.} \bibinfo{year}{2022}\natexlab{}.
\newblock \showarticletitle{{AMD} Instinct {MI200} Series Accelerator and Node Architectures}. In \bibinfo{booktitle}{\emph{2022 IEEE Hot Chips 34 Symposium (HCS)}}. IEEE Computer Society, \bibinfo{pages}{1--23}.
\newblock


\bibitem[Top500(2023)]%
        {top500}
\bibfield{author}{\bibinfo{person}{Top500}.} \bibinfo{year}{2023}\natexlab{}.
\newblock \bibinfo{title}{November 2023}.
\newblock
\newblock
\urldef\tempurl%
\url{https://top500.org/lists/top500/2023/11/}
\showURL{%
\tempurl}


\bibitem[{Unified Acceleration Foundation}(2023)]%
        {oneAPI}
\bibfield{author}{\bibinfo{person}{{Unified Acceleration Foundation}}.} \bibinfo{year}{2023}\natexlab{}.
\newblock \bibinfo{title}{{oneAPI}}.
\newblock
\newblock
\urldef\tempurl%
\url{https://www.oneapi.io/spec/}
\showURL{%
\tempurl}


\bibitem[Williams et~al\mbox{.}(2009)]%
        {williams2009roofline}
\bibfield{author}{\bibinfo{person}{Samuel Williams}, \bibinfo{person}{Andrew Waterman}, {and} \bibinfo{person}{David Patterson}.} \bibinfo{year}{2009}\natexlab{}.
\newblock \showarticletitle{Roofline: an insightful visual performance model for multicore architectures}.
\newblock \bibinfo{journal}{\emph{Commun. ACM}} \bibinfo{volume}{52}, \bibinfo{number}{4} (\bibinfo{year}{2009}), \bibinfo{pages}{65--76}.
\newblock


\bibitem[Winter et~al\mbox{.}(2018)]%
        {faimgraph}
\bibfield{author}{\bibinfo{person}{Martin Winter}, \bibinfo{person}{Daniel Mlakar}, \bibinfo{person}{Rhaleb Zayer}, {et~al\mbox{.}}} \bibinfo{year}{2018}\natexlab{}.
\newblock \showarticletitle{faimGraph: High Performance Management of Fully-Dynamic Graphs Under Tight Memory Constraints on the GPU}. In \bibinfo{booktitle}{\emph{SC18: International Conference for High Performance Computing, Networking, Storage and Analysis}}. \bibinfo{pages}{754--766}.
\newblock
\urldef\tempurl%
\url{https://doi.org/10.1109/SC.2018.00063}
\showDOI{\tempurl}


\bibitem[Xu et~al\mbox{.}(2022)]%
        {xuUVMhints}
\bibfield{author}{\bibinfo{person}{Hailu Xu}, \bibinfo{person}{Pei-Hung Lin}, \bibinfo{person}{Murali Emani}, {et~al\mbox{.}}} \bibinfo{year}{2022}\natexlab{}.
\newblock \showarticletitle{XUnified: A Framework for Guiding Optimal Use of {GPU} Unified Memory}.
\newblock \bibinfo{journal}{\emph{IEEE Access}}  \bibinfo{volume}{10} (\bibinfo{year}{2022}), \bibinfo{pages}{82614--82625}.
\newblock


\bibitem[Yu et~al\mbox{.}(2019)]%
        {Yu2019b}
\bibfield{author}{\bibinfo{person}{Qi Yu}, \bibinfo{person}{Bruce Childers}, \bibinfo{person}{Libo Huang}, {et~al\mbox{.}}} \bibinfo{year}{2019}\natexlab{}.
\newblock \showarticletitle{A quantitative evaluation of unified memory in {GPUs}}.
\newblock \bibinfo{journal}{\emph{The Journal of Supercomputing}} \bibinfo{volume}{76}, \bibinfo{number}{4} (\bibinfo{date}{nov} \bibinfo{year}{2019}), \bibinfo{pages}{2958--2985}.
\newblock


\end{thebibliography}

\end{document}